\documentclass[a4paper]{article}
\usepackage[english]{babel}
\usepackage[utf8x]{inputenc}
\usepackage[T1]{fontenc}

\usepackage[a4paper,top=3cm,bottom=2cm,left=3cm,right=3cm,marginparwidth=1.75cm]{geometry}

\usepackage{amsmath}
\usepackage{graphicx}
\usepackage{caption}
\usepackage{subcaption}
\usepackage{adjustbox}
\usepackage{booktabs}
\usepackage{epstopdf}
\usepackage{placeins}
\usepackage{float}
\usepackage{multirow}
\usepackage[colorlinks=true, allcolors=blue]{hyperref}
\usepackage{cite}

\usepackage{setspace}
\title{Particle contact dynamics as the origin for noninteger power expansion rheology in attractive suspension networks}
\author{Irene Natalia$^1$, Randy H. Ewoldt$^2$, Erin Koos$^{1,\ast}$}
 \date{ \small
 $^1$ {KU Leuven, Soft Matter, Rheology and Technology - Department of Chemical Engineering, Celestijnenlaan 200f, 3001 Leuven, Belgium} \\
$^2$ {Department of Mechanical Science and Engineering, University of Illinois at Urbana-Champaign, Urbana, IL 61801, USA} \\
 $^{\ast}$ E-mail: erin.koos@kuleuven.be \\}

\begin{document}

%\tracingall
\maketitle

\begin{abstract}
We show that Hertzian particle contacts are the underlying cause of the as-yet-unexplained noninteger power laws in weakly nonlinear rheology.
In the medium amplitude oscillatory shear (MAOS) region, the cubic scaling of the leading order nonlinear shear stress ($\sigma_\mathrm{3} \sim \gamma_\mathrm{0}^{m_\mathrm{3}}$, $m_\mathrm{3}=3$) is the standard expectation. Expanding on the work by Natalia {\it et al.} [J. Rheol. {\bf 64} 625--635 (2020)], we report an extensive data set of noncubical, noninteger power law scalings $m_\mathrm{3}$ for particle suspensions in two immiscible fluids with a capillary attractive interaction, known as capillary suspensions. Here, we show that distinct power law exponents are found for the storage and loss moduli and these noninteger scalings occur at every secondary fluid concentration for two different contact angles. These compelling results indicate that the noninteger scalings are related to the underlying microstructure of capillary suspensions. We show that the magnitude of the third harmonic elastic stress scaling $m_\mathrm{3,elastic}$ originates from Hertzian-like contacts in combination with the attractive capillary force. The related third harmonic viscous stress scaling $m_\mathrm{3,viscous}$ is, found to be associated with adhesive-controlled friction. These observations, conducted for a wide range of compositions, can help explain previous reports of noninteger scaling for materials involving particle contacts and offers a new opportunity using the variable power law exponent of MAOS rheology to reveal the physics of particle bonds and friction in the rheological response under low deformation instead of at very high shear rates.
\end{abstract}

\section{Introduction} \label{introduction}
Medium amplitude oscillatory shear (MAOS) is an advanced characterization technique that can provide deep insight into material structures and interactions, e.g.~it was used to resolve a nearly 70-year debate concerning which molecular processes cause dramatic elastic-stiffening and viscous-thickening of a canonical transient polymer network~\cite{Martinetti.2018.Macromol}. Although the theoretical paradigm of MAOS has existed for many decades~\cite{Kirkwood.1956}, recent experimental observations contradict basic underlying assumptions. Particularly, noninteger power law scalings of stress versus strain have been observed in particle-based systems~\cite{Natalia.2020}. This raises concerns about the validity of the MAOS technique, given the current absence of a physical explanation. 

In this asymptotically nonlinear MAOS regime, the shear stress response becomes nonlinear with the appearance of a third harmonic and nonlinearity of the first harmonic, but all higher order harmonics are negligibly small~\cite{Ewoldt.2013}. The application of an increasing strain amplitude $\gamma_\mathrm{0}$ usually results in a scaling of $\sigma_\mathrm{3} \sim \gamma_\mathrm{0}^{3}$ for the deviation from linearity \cite{Ewoldt.2013, Wagner.2011, Reinheimer.2012, Bharadwaj.2014b, Hyun.2009, Swan.2016}. However, a noncubic and indeed noninteger $m_\mathrm{3}$ in the scaling $\sigma_\mathrm{3} \sim \gamma_\mathrm{0}^{m_\mathrm{3}}$ has been reported for some materials \cite{Kadar.2017, Nam.2011,Hyun.2007, Blackwell.2016, Hirschberg.2020}. This was most conclusively demonstrated to be a property of the material tested rather than an instrumental artifact in our previous work \cite{Natalia.2020}. Moreover, a comparison of the materials exhibiting this noninteger scaling points to particle contacts as a common feature in nearly all reported observations of noninteger and noncubic MAOS scalings, thus, a potential origin of this peculiar scaling. Particle contacts may be strongly nonlinear due to Hertzian contact mechanics, frictional contact mechanics, or a combination thereof and are, therefore a good candidate for the observed anomalous scaling. However, relating these effects to weakly nonlinear MAOS rheology has not previously been made.

Direct Hertzian contacts between colliding particles have been reported as the cause of discontinuous shear thickening (DST) in dense suspensions, shifting the jamming point to a lower critical volume fraction compared to a frictionless system \cite{Mari.2014}. DST is observed for colloidal and noncolloidal suspensions at high shear rates, in the regime where frictional interactions dominate \cite{Seto.2013, Wyart.2014, Mari.2015}. These Hertzian contacts are inherently nonlinear and this nonlinearity was recently shown to play an important role in the the shear thinning of concentrated suspensions and the critical jamming volume fraction~\cite{Lobry.2019, More.2020}. However, the importance of friction and particle contacts, especially their influence on the rheological response at small deformation amplitudes, e.g. in the asymptotically nonlinear regime, is still largely unexplained. 
Since thermal motion causes diffusive particle motion and keeps particles well distributed, noncolloidal suspensions are an interesting system to study the hydrodynamic effect on the suspension stress without the complication of Brownian motion. The repulsive contribution from a stabilization layer in the noncolloidal suspensions can often be neglected, which consequently leads to particle-particle contacts even under slight hydrodynamic influence, or in other words, at very small deformation or shear rates \cite{Mewis.2012}. 

The addition of a small amount of immiscible fluid to the bulk phase of a particle suspension causes a capillary attractive interaction and induces a percolated sample-spanning network \cite{Koos.2011}. Thus, this kind of suspension is called a capillary suspension. This effect results in an increase of yield stress and occurs independently from the wettability of the secondary fluid to the particles. Suspensions that are formed with a better wetting secondary fluid are called capillary suspensions in the \emph{pendular state}; the contact angle of the ternary system is small and the pendular bridges have concave menisci. On the other hand, capillary suspensions with the bulk fluid as the better wetting liquid are in the \emph{capillary state} and the liquid bridges have a convex meniscus. The particles in the capillary state will aggregate around small secondary fluid droplets to minimize energy, which will result in a short-range attractive force \cite{Koos.2012}. A sketch of both states of capillary suspensions is shown in Figure \ref{fig:ternary_system}.

Capillary suspensions offer a novel, yet simple route to tune the rheological properties of the material with many potential applications, such as low-fat spreadable chocolate \cite{Wollgarten.2015}, printable electronics \cite{Schneider.2016}, reduction of cracks in thin films \cite{Schneider.2017}, precursors for porous ceramic or glass or polymer membranes \cite{Dittmann.2012, Maurath.2015, Hauf.2017}. It is imperative to understand how these suspensions behave under shear for them to reach their full potential. From the general rheological perspective, the presence of a percolated network, even at low solid volume fractions, makes this system an interesting model to study the influence of hydrodynamic and contact forces on the rheological properties. Unlike shear thickening materials, the particles in the capillary suspensions are already in contact or close to contact at quiescence due to the attractive capillary force, allowing us to study the effect of particle contacts even at small deformations. Furthermore, capillary suspensions offer unique capabilities to tune the particle contacts, making them an ideal system to study the role of the these contacts in the atypical noninteger MAOS scaling.

In the present paper, we employ MAOS experiments to understand what happens with the particle bonds in capillary suspensions at small deformations. In the previous work, we reported that the third harmonic elastic and viscous stresses scaled in an atypical noncubical, noninteger manner with the strain for one specific composition of capillary suspension in the capillary state \cite{Natalia.2020}. Although MAOS stress output consists of four signals of weak nonlinearity:  the third harmonics $\sigma'_\mathrm{3}$ and $\sigma''_\mathrm{3}$, as well as the deviation of the first harmonics from their linear value $\sigma'_1-G'_\mathrm{LVE} \cdot \gamma_0$ and $\sigma''_1-G''_\mathrm{LVE} \cdot \gamma_0$, our focus will be on the third harmonic stress signals since they have less uncertainty above the noise floor. Here, we report and discuss the third harmonic of elastic and viscous suspension stress produced by MAOS for capillary suspensions with different concentrations of secondary fluid in both the pendular and capillary states. We will show that all formulations tested exhibit noncubical, noninteger scaling, and then propose a Hertzian contact model to rationalize this response.

%%%%  Section Material and Methods %%%%%
\section{Materials and methods} \label{material_method}

\subsection{Capillary suspensions} \label{CapS_preparation}

We used two different model systems for the ternary particle-liquid-liquid suspensions, one in the pendular and one in the capillary state. The first system consists of NP3 silica glass particles (3.55~{$\mathrm{\mu}$}m diameter) in silicone oil with added glycerol, where the glycerol is the better wetting liquid. The second model system consists of PMMA particles (22.5~{$\mathrm{\mu}$}m diameter) in glycerol with added paraffin oil, with the paraffin oil as the less wetting fluid. The RMS (root mean square) roughness of the NP3 silica glass was measured as 3.3~nm using AFM (Bruker Multimode 8 with a Nanoscope V controller) using an OMCL-AC160TS-R3 probe in tapping mode with a spring constant of approximately 20~N/m and a resonance frequency of approximately 300~kHz. A scan rate of 0.5--1~Hz was used. The RMS roughness was obtained using the Gwyddion software using images of 1 by 1~{$\mathrm{\mu}$}m$^2$. The PMMA beads, produced via an emulsion route, have a roughness of 2.2~nm, although there are also some regions with adsorbed nanoparticles (RMS roughness of 27~nm).

\begin{figure}[ht]
    \centering
    \includegraphics[width=0.5\textwidth]{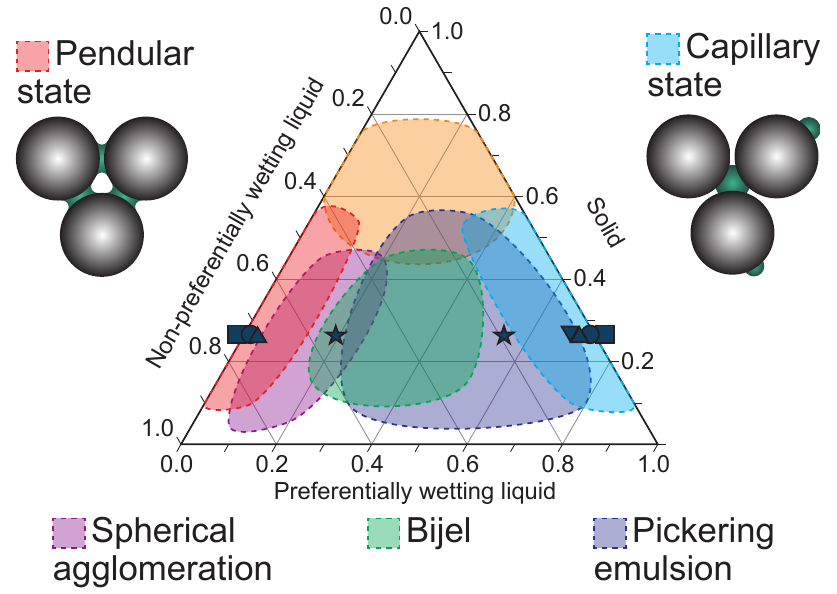}
		\caption{Schematic of ternary solid-liquid-liquid systems used in this work. Adapted from Koos~\cite{Koos.2014}. The solid volume fraction is constant $(\phi_\mathrm{solid}=0.25)$. The symbols on the left side represent the NP3-silicone oil-glycerol samples with a different concentration of the secondary fluid $(\phi_\mathrm{2nd}/\phi_\mathrm{solid}=0.0-0.8)$, accessing the pendular state. The PMMA-glycerol-paraffin oil samples with various concentrations of paraffin oil are depicted on the right side, accessing the capillary suspensions in the capillary state $(\phi_\mathrm{2nd}/\phi_\mathrm{solid}=0.0-0.8)$.}%
    \label{fig:ternary_system}
\end{figure}

The NP3-silicone oil-glycerol model system represents the symbols on the left side of Figure \ref{fig:ternary_system} that access the pendular state and the PMMA-glycerol-paraffin oil represents the symbols on the right side of the schematic that access the capillary state. For all samples, we kept the solid concentration constant at $\phi_\mathrm{solid}=0.25$. We varied the secondary fluid concentration so the samples cover the normal suspension ($\phi_\mathrm{2nd}/\phi_\mathrm{solid}=0.0$) and capillary suspensions in the pendular state ($\phi_\mathrm{2nd}/\phi_\mathrm{solid}=0.05$, 0.2) and the bicontinuous state ($\phi_\mathrm{2nd}/\phi_\mathrm{solid}=0.8$) for the NP3-silicone oil-glycerol. Analogously, PMMA-glycerol-paraffin oil samples cover the transition from a normal suspension ($\phi_\mathrm{2nd}/\phi_\mathrm{solid}=0.0$) to capillary suspensions in the capillary state ($\phi_\mathrm{2nd}/\phi_\mathrm{solid}=0.1$ -- 0.3) and Pickering-emulsion-like state ($\phi_\mathrm{2nd}/\phi_\mathrm{solid}=0.8$). We use the term Pickering-emulsion-like state as the size of the particles and the secondary fluid droplets in this paper can be of the same order, unlike the typical case of a Pickering emulsion where small particles stabilize the interface of a large drop.
The details of each sample used in this work are given in Table \ref{table:sample_composition}. The method of sample preparation for each model system was described in the study by Natalia {\it et al.} \cite{Natalia.2018}.

\begin{table}[ht]
\normalsize
\caption{Overview of the sample compositions used in this manuscript} 
\centering 
\begin{adjustbox}{width=\textwidth}
\begin{tabular}{lp{0.25\textwidth}p{0.25\textwidth}p{0.25\textwidth}p{0.1\textwidth}*{2}{l}}
\toprule
State & Solid & Bulk fluid & Secondary fluid & Contact angle $\theta$ & $\phi_\mathrm{solid}$ & $\phi_\mathrm{2nd}/\phi_\mathrm{solid}$\\
\midrule
Pendular & Silica oxide glass & Silicone oil & Glycerin & $70\pm7^{\circ}$ & 0.25 & 0, 0.05, 0.2, 0.8\\
 & (OMicron NP3-P0) & (Wacker AK 200) & (Rotipuran $\geq 99.5\%$) & & & \\
 & Sovitec, Fleurus, Belgium & Wacker Chemie AG, Munich, Germany & Carl Roth, Karlsruhe, Germany & & & \\
 & d$_{50,3}=3.55 \pm 0.04$~{$\mathrm{\mu}$}m & $\eta=0.2$ Pa$\cdot$s & $\eta=1.412$ Pa$\cdot$s & &\\
 & $\rho_\mathrm{(20^{\circ}C)}=2.46$ g/ml & $\rho_\mathrm{(20^{\circ}C)}=1.07$ g/ml & $\rho_\mathrm{(20^{\circ}C)}=1.26$ g/ml & & & \\
\\
Capillary & PMMA beads & Glycerin & Paraffin oil & $88\pm4^{\circ}$ & 0.25 & 0, 0.1, 0.2, 0.3, 0.8\\
 & (Altuglas BS100) & (Rotipuran $\geq 99.5\%$) & & & & \\
 & Altuglas International, La Garenne-Colombes, France & Carl Roth, Karlsruhe, Germany & Sigma-Aldrich Chemie GmbH, Steinheim, Germany & & &\\
 & d$_{50,3}=22.5\pm 0.06$~{$\mathrm{\mu}$}m & $\eta=1.412$ Pa$\cdot$s & $\eta=0.21$ Pa$\cdot$s & & & \\
 & $\rho_\mathrm{(20^{\circ}C)}=1.2$ g/ml & $\rho_\mathrm{(20^{\circ}C)}=1.26$ g/ml & $\rho_\mathrm{(20^{\circ}C)}=0.88$ g/ml & & & \\
\bottomrule
\end{tabular}
\end{adjustbox}
\label{table:sample_composition} 
\end{table}

\subsection{Rheological characterization} \label{rheological_characterization}

All rheological measurements were conducted with a TA Instruments ARES-G2 rotational rheometer (separated motor-transducer) using the titanium plate-plate geometry with 50 mm diameter and at 1 mm gap height. Two types of sandpaper were used to eliminate wall slip for the pendular state samples: P320 (grit size = 46.2~{$\mathrm{\mu}$}m) for $\phi_\mathrm{2nd}/\phi_\mathrm{solid}\leq 0.05$ and P80 (grit size = 201~{$\mathrm{\mu}$}m) for $\phi_\mathrm{2nd}/\phi_\mathrm{solid}\geq 0.2$.  Flow sweep measurements at different gap heights were executed as preliminary tests for the capillary state samples to ensure that no slip occurs. All tests were performed at $20^{\circ}$C and all reported measurements were executed at least three times to check their reproducibility. The multiple strain-amplitude sweeps protocol is given in detail in Natalia {\it et al.} \cite{Natalia.2020}.

We conducted multiple strain-amplitude sweeps for $0.01\% \leq \gamma_\mathrm{0} \leq 1000\%$  with four arbitrary maximum amplitudes ($\gamma_\mathrm{0,max}=$1, 10, 100, and 1000$\%$) at a constant frequency $\omega=0.628$ rad/s in the correlation mode, sweeping from a low amplitude ($\gamma_\mathrm{0,min}=0.01\%$) to the corresponding maximum amplitude and back from high to low amplitude without delay. This allowed us to ensure that the measured nonlinear response was reversible. Each experiment subset with a specific $\gamma_\mathrm{0,max}$ was run three consecutive times to ensure that our result is not caused by time effects or evaporation. This additionally precluded an irreversible evolution of the sample structure. We started with a maximum amplitude of $\gamma_\mathrm{0,max}=1\%$ and subsequently increased $\gamma_\mathrm{0,max}$ to the next higher value ($10\%$) using the same sample after each strain-amplitude sweep (forward and reverse order) was performed three times. This measurement set, consisting of 12 forward and 12 reverse amplitude sweeps, was repeated using three different samples to test the repeatability. While complex, the measurement protocol allows us to ensure that the data are not affected by experimental error, we are not measuring instrument or sample noise, the system is reversible, and we are only fitting the MAOS regime~\cite{Natalia.2020}.

\begin{figure}[tb]
    \centering
    \includegraphics[width=\textwidth]{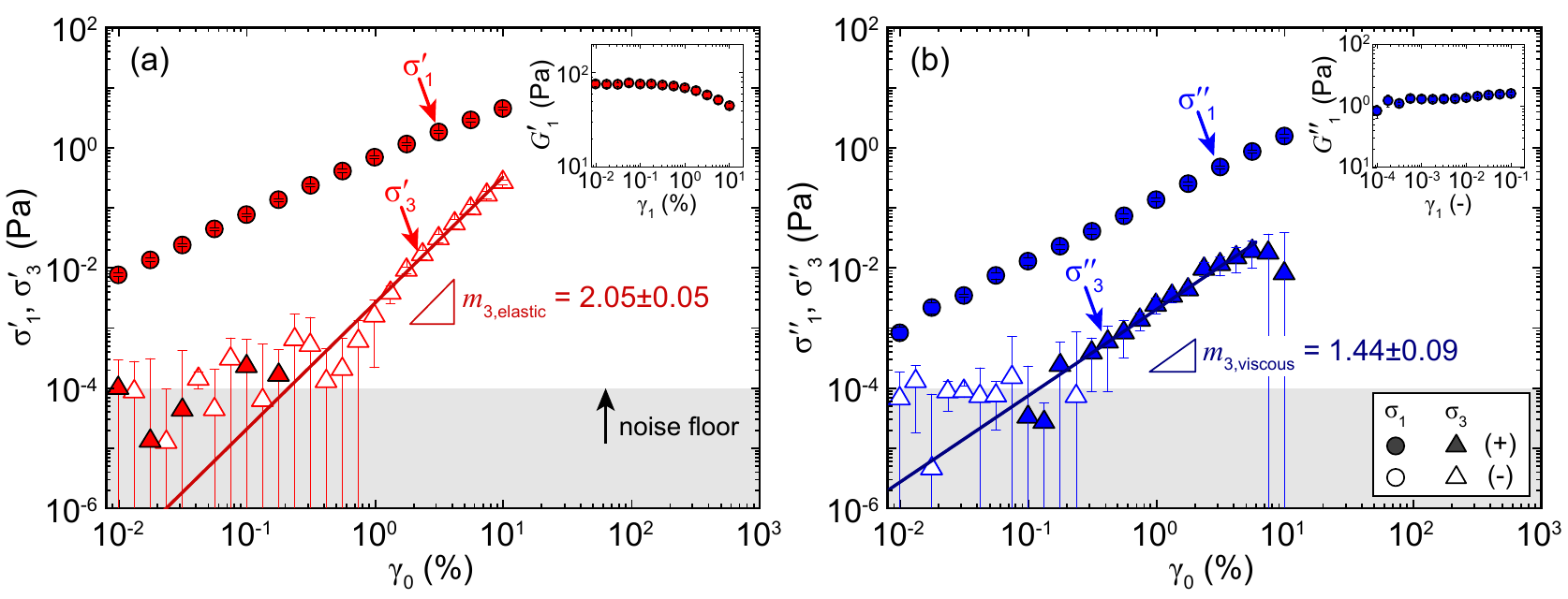}
		\caption{Calculation of the two power law scalings ($m_\mathrm{3,elastic}$ and $m_\mathrm{3,viscous}$) from the third harmonic stresses, as measured via a strain-amplitude sweep (capillary state sample with $\phi_\mathrm{2nd}/\phi_\mathrm{solid}=0.1$ at $\gamma_\mathrm{0,max}=10\%$, $2^\mathrm{nd}$ up). Open symbols denote negative values and filled symbols positive values. (a) First $(\sigma'_1)$ and third $(\sigma'_3)$ elastic stress, where the inset shows the elastic modulus $G'_1=\sigma'_1 / \gamma_0$. (b) First $(\sigma''_1)$ and third $(\sigma''_3)$ viscous stress with inset showing the viscous modulus $G''_1=\sigma''_1 / \gamma_0$. Figure modified from \cite{Natalia.2020}, reprinted with permission.}%
    \label{fig:ExampleMAOS}
\end{figure}

An example result from a single strain-amplitude sweep is shown in Figure \ref{fig:ExampleMAOS} for a capillary state sample with $\phi_\mathrm{2nd}/\phi_\mathrm{solid}=0.1$ at $\gamma_\mathrm{0,max}=10\%$ ($2^\mathrm{nd}$ up measurement). 
The asymptotic nonlinearities are calculated from the four stress coefficients of the Fourier series representation of the stress response~\cite{Ewoldt.2013b}, $\sigma'_\mathrm{1}$, $\sigma''_\mathrm{1}$, $\sigma'_\mathrm{3}$ and $\sigma''_\mathrm{3}$. The first harmonics are fit using an asymptotic expansion,
\begin{equation}
		\sigma'_\mathrm{1}(\omega,\gamma_\mathrm{0})=G'_\mathrm{LVE}(\omega) \cdot \gamma_\mathrm{0}+[e_\mathrm{1}](\omega)\cdot\gamma_\mathrm{0}^{m_\mathrm{1,elastic}}+\mathcal{O}(\gamma_\mathrm{0}^{p_\mathrm{1,elastic}}) ,
		%\label{eq:sigma_prime_1_deviation}
\end{equation}
\begin{equation}
		\sigma''_\mathrm{1}(\omega,\gamma_\mathrm{0})=G''_\mathrm{LVE}(\omega) \cdot \gamma_\mathrm{0}+\omega [v_\mathrm{1}](\omega) \cdot \gamma_\mathrm{0}^{m_\mathrm{1,viscous}}+\mathcal{O}(\gamma_\mathrm{0}^{p_\mathrm{1,viscous}}) .
		%\label{eq:sigma_doubleprime_1_deviation}
\end{equation}
The nonlinearities in the third harmonic are given by
\begin{equation}
			\sigma'_\mathrm{3}(\omega,\gamma_\mathrm{0})=-[e_\mathrm{3}](\omega)\cdot \gamma_\mathrm{0}^{m_\mathrm{3,elastic}}+\mathcal{O}(\gamma_\mathrm{0}^{p_\mathrm{3,elastic}}) ,
			\label{eq:sigma_prime_3}
\end{equation}
\begin{equation}
			\sigma''_\mathrm{3}(\omega,\gamma_\mathrm{0})=\omega[v_\mathrm{3}](\omega) \cdot \gamma_\mathrm{0}^{m_\mathrm{3,viscous}}+\mathcal{O}(\gamma_\mathrm{0}^{p_\mathrm{3,viscous}}) ,
			\label{eq:sigma_doubleprime_3}
\end{equation}
where $[e_\mathrm{1}](\omega),[v_\mathrm{1}](\omega), [e_\mathrm{3}](\omega)$, and $[v_\mathrm{3}](\omega)$ are the four intrinsic nonlinear material functions~\cite{Ewoldt.2013}. The letter ``$e$'' denotes elastic nonlinearities and ``$v$'' is for viscous nonlinearities.
In the present paper, we report only the power law fits obtained from the third harmonics since they are cleaner and have less uncertainty above the minimum stress in the oscillation mode ($\sigma_\mathrm{min} = 10^{-4}$~Pa, which is chosen from the power spectrum of the stress harmonics and corresponds to a minimum torque of $\mathrm{T_{min}}\approx 5$~nN$\cdot$m).

To determine the power law scalings $m_\mathrm{3,elastic}$ and $m_\mathrm{3,viscous}$, we used a fitting procedure that weighted the points by their uncertainty. This weight used the maximum of the standard deviation from triplicate measurements and lower stress limit. Since the scaling of the asymptotic nonlinearity is also very sensitive to the upper fitting range, we chose a fit range that minimized the error in the slope for the various possibilities in the number of data points.  The fitting procedure and an extended discussion about the measurement certainty is described in the study by Natalia {\it et al.} \cite{Natalia.2020}. Within current experimental limits, we are confident that our fitting procedure captures the most credible power law apparent in the data.

%%%%  Section Hertzian Contact Model %%%%%
\section{Hertzian contact model} \label{Hertzian_model}

\subsection{Contact between two ideal spheres}

\begin{figure}[h]
	\centering
			\includegraphics[width=0.5\textwidth]{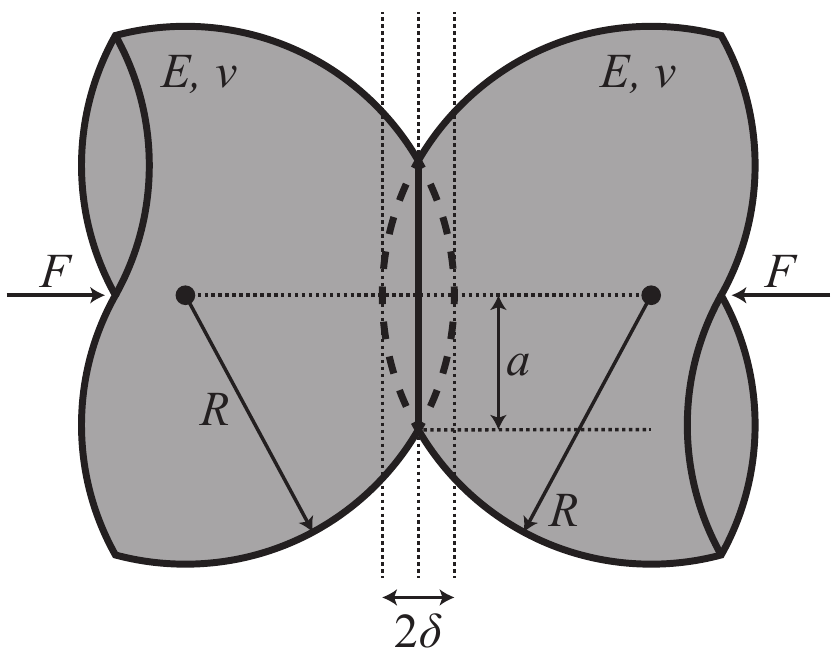}
	\caption[Schematic of Hertzian contact between two ``soft'' particles.]
	{Schematic of the Hertzian contact between two identical elastic spheres with a radius $R$, Young's modulus $E$, and Poisson ratio $\nu$. The penetration depth of each sphere is $\delta$ and $a$ is the contact radius.}
	\label{fig:hertzian_contact}
\end{figure}

Upon a particle-particle contact, the elastic particle contact force $F$ can be described using the linear Hookean relationship ($F\sim \delta$) with indentation depth $\delta$ only if the contact area between the two bodies is constant. However, the nonlinear Hertzian relationship ($F\sim \delta^{3/2}$) is used if the contact area between two linearly elastic spherical bodies continuously changes, as is the case for two deformable particles. The Hertzian contact theory is often applied to granular materials where particles have a well-defined diameter and do not interact except for this strong repulsive force, which limits the particle deformation \cite{Coste.1999, OHern.2003, Corwin.2005, Owens.2011}. In dispersions, the Hertzian contact is commonly used to describe the contact between soft particles, such as microgels or deformable emulsions \cite{Seth.2011, Ghosh.2019, Khabaz.2020}, but such interactions might also influence the rheological response of boehmite nanoparticle suspensions \cite{Finke.2020}.
Using the Hertzian contact model, the elastic repulsion force is given by \cite{Johnson.1985} 
\begin{equation}
	F=
		\begin{cases}
			\frac{4}{3}E^*{R^*}^{1/2}\delta^{3/2} & \text{for $\delta>0$ , }\\
			0 & \text{for $\delta\leq0$ , }
		\end{cases}		
\label{eq: Hertzian_force}
\end{equation}
with
\begin{equation}
	\frac{1}{E^*}=\frac{1-\nu_\mathrm{1}^2}{E_\mathrm{1}}+\frac{1-\nu_\mathrm{2}^2}{E_\mathrm{2}},
\label{eq:E_definition}
\end{equation}
and
\begin{equation}
	R^*=\frac{R_\mathrm{1}R_\mathrm{2}}{R_\mathrm{1}+R_\mathrm{2}},
\label{eq:eff_R_definition}
\end{equation}
where $R^*$ is the effective particle radius, calculated from the radii of two different spheres $R_\mathrm{1}$ and $R_\mathrm{2}$, $E^*$ is the effective Young's modulus, determined from $E_\mathrm{1}$ and $E_\mathrm{2}$, the Young's modulus of the two different spheres, and their Poisson ratios $\nu_\mathrm{1}$ and $\nu_\mathrm{2}$. In case of two identical spheres with the same $R,E$, and $\nu$, equations \ref{eq:E_definition} and \ref{eq:eff_R_definition} reduce to $E^*=\frac{E}{2\left(1-\nu^2\right)}$ and $R^*=R/2$. Furthermore, for two elastic spheres in contact, $2\delta$ can be defined as the reduction in center-to-center distance as displayed in Figure \ref{fig:hertzian_contact}. With increasing displacement or penetration depth $\delta$, the contact radius $a$ changes by 
\begin{equation}
 a^2=\delta R^*.
\label{eq: contact_radius_a}
\end{equation}
This model is valid in the limit of small strain deformation, where $\delta\ll R^*$. A schematic of a Hertzian contact is shown in Figure \ref{fig:hertzian_contact}. Ideal hard-sphere particles with hard, undeformable surfaces exhibit $E=\infty$, and, therefore, any real sphere can be considered as a ``soft'' elastic particle.

\begin{figure*}[h]%
    \centering
    \includegraphics[width=\textwidth]{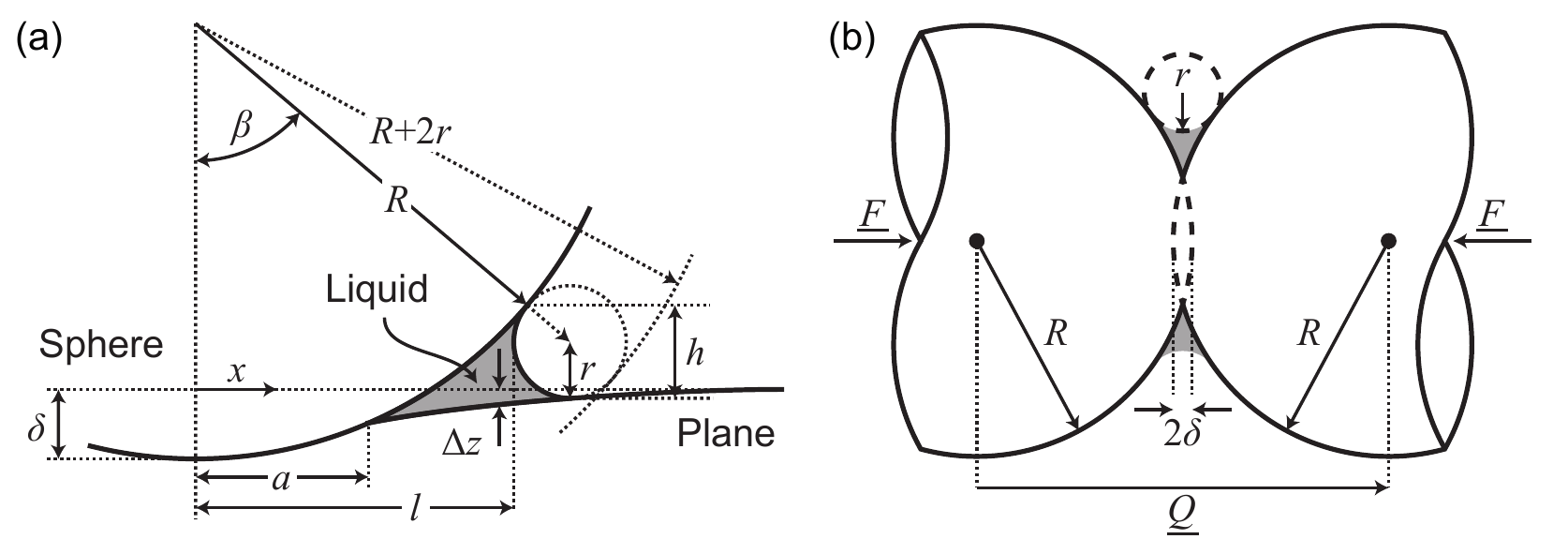}
	\caption[Schematic of the Hertzian contact with adhesive capillary force.]{Schematic of Hertzian contact force in combination with adhesive capillary force. (a)~Schematic of the elastically deformed surface, where $a$ is the contact radius, $\beta$ is the filling angle, $r$ and $l$ are the the principal radii of curvature of the liquid bridge, and $h$ is the height of liquid bridge. (b)~Two `soft' particles experience an additional deformation due to the presence of a liquid bridge. $\underline{Q}$ is a vector describing the center-to-center particle distance, $R$ is the particle radius and $\delta$ is the indentation depth of each sphere. Subfigure (a) adapted with permission from The Royal Society of Chemistry \cite{Butt.2010}.}%
\label{fig:force_capillary}%
\end{figure*}

The interaction force $F$ can arise from external boundary conditions or attractive interactions between particles. In the presence of, e.g., a van der Waals (vdW) force, this attractive surface force pulls the two surfaces together, causing a contact area even under zero external load \cite{Israelachvili.2011}. 
The strong capillary force, the attractive potential of which can be $\sim 1000$ kT \cite{Danov.2001}, greatly amplifies this preloading of the contact. For the silica system used here, we calculate a force $F_c \approx 1.8 \times 10^{-10}$~N and nominal stress $F_c/(\pi a^2) \approx 20$~Pa. While this is on the same order of the experimentally applied shear stresses, such a simplification ignores the properties of the multiply connected capillary network. In reality, these systems exhibit a yield stress and very little deformation under shear. A strong attractive force causes the deformation and flattening of the particles \cite{Johnson.1971}, which modifies the principal radius of curvature $r$. This leads to a pull-off force required to detach the particles from contact in the presence of a liquid meniscus \cite{Butt.2010}, as shown in Figure \ref{fig:force_capillary}a) 
\begin{equation}
		F=-\frac{4}{3}E^* {R^*}^{1/2}\delta^{3/2}+4\pi\Gamma R^*\left(1+\frac{\delta}{4r}\right).
	\label{eq:Hertzian_capillary_force}
\end{equation}
The first term on the right-hand side describes the repulsive `compressive' Hertzian contact force and the next two terms depict the contribution of the attractive `tensile' capillary force, given by the principal radii of curvature $r$ and $l$ of a toroidal, liquid bridge (Figure \ref{fig:force_capillary}a). For the capillary force term, it is assumed that the ternary contact angle $\theta=0^{\circ}$ and $R\gg l \gg r$; therefore, the line traction contribution is negligible compared to the Laplace pressure effect. A detailed derivation of the equation can be found in \cite{Pietsch.1967, Butt.2010}.

\subsection{Implications for MAOS measurements}

We hypothesize that the Hertzian contact force, initiated by the adhesive capillary force, is the origin of the noninteger stress scaling; thus, it is important to transform the function of $F(\delta)$ into $\sigma(\gamma_\mathrm{0})$. Using the Kramers expression, the relation between microscopic force to the macroscopic stress tensor is described as \cite{Bird.1987b}
\begin{equation}
	\boldsymbol{\sigma}=n \left\langle \underline{F} \underline{Q}\right\rangle
\label{eq:Kramers_stress_calculator}
\end{equation}
where $n$ describes the number of the force elements per volume, $\underline{F}$ is the tensile force in the force element, as depicted in equation \ref{eq:Hertzian_capillary_force}, and $\underline{Q}$ is the center-to-center particle vector over which the $\underline{F}$ acts (Figure \ref{fig:force_capillary}b). It is assumed that both particles have an identical radius $R$, thus $\underline{Q}$ is defined geometrically by
\begin{equation}
	\underline{Q}=2(R-\delta) \underline{e}_Q
\label{eq:definition_Q}
\end{equation}
where $\underline{e}_Q$ denotes the unit vector in the direction of $\underline{Q}$. Using the absolute distance between the particle centers $\left|\underline{Q}\right|=Q$, 
Equation \ref{eq:definition_Q} can be rearranged into
\begin{equation}
	\delta=R-\frac{1}{2}\frac{\underline{Q}}{\underline{e}_Q}=R-\frac{1}{2}Q.
\label{eq:definition_delta}
\end{equation}
If it is assumed that both particles are identical (identical radius $R$, Young's modulus $E$, and Poisson ratio $\nu$), a substitution of equation \ref{eq:definition_delta} into equation \ref{eq:Hertzian_capillary_force} results in
\begin{equation}
	%\underline{F}= -\frac{4}{3}E^*R^{1/2}\left(R-\frac{1}{2}|Q|\right)^{3/2}\cdot \underline{e}_Q+2\pi\Gamma R\left(1+\frac{\left(R-\frac{1}{2}|Q|\right)}{4r}\right)\cdot \underline{e}_Q
	\underline{F}= -\frac{4}{3}E^*\left(\frac{1}{2}R\right)^{1/2}\left(R-\frac{1}{2}Q\right)^{3/2} \underline{e}_Q+2\pi\Gamma R\left[1+\frac{\left(R-\frac{1}{2}Q\right)}{4r}\right] \underline{e}_Q
\label{eq:F_function_Q}
\end{equation}
Thus, the stress tensor $\boldsymbol{\sigma}$ can be calculated as
\begin{equation}
	%\boldsymbol{\sigma}=n\left\langle \left(-\frac{4}{3}E^*R^{1/2}\left(R-\frac{1}{2}Q\right)^{3/2} + 2\pi\Gamma R +  \frac{\pi\Gamma R}{2r}\left(R-\frac{1}{2}Q\right)  \right) Q \cdot \underline{e}_Q \cdot \underline{e}_Q \right\rangle
	\boldsymbol{\sigma}=n\left\langle \left[-\frac{2\sqrt{2}}{3}E^*R^{1/2}\left(R-\frac{1}{2}Q\right)^{3/2} + 2\pi\Gamma R +  \frac{\pi\Gamma R}{2r}\left(R-\frac{1}{2}Q\right)  \right] Q \underline{e}_Q  \underline{e}_Q \right\rangle
\label{eq:stress_function_Q}
\end{equation}
where the angle brackets represent the ensemble average. From equation \ref{eq:stress_function_Q}, it is clear that the stress tensor depends on the distance between two particle centers $Q=\left|\underline{Q}\right|$. At some nonzero $\delta=\delta_\mathrm{equil}$, the Hertzian repulsion is balanced by the capillary attraction, and, therefore, the stress will be zero.  Under shear deformation, the variable $Q$ can be linked to the shear strain $\gamma$ or shear rate $\dot{\gamma}$.
A detailed mathematical derivation to show the correlation between the dyadic product of $\left\langle Q Q \underline{e}_Q  \underline{e}_Q \right\rangle$ and $\gamma$ is given in the study by Bharadwaj {\it et al.} \cite{Bharadwaj.2017}. Assuming a limit of high Deborah number, e.g. $\mathrm{De}=\lambda \cdot \omega > 62.8$ based on the experimentally applied oscillatory frequency $\omega=0.628$~rad/s and the characteristic relaxation time of the material $\lambda>100$~s, as well as affine deformation gives $\left\langle Q_2 Q_1 \right\rangle = \gamma_\mathrm{0}$. From equation \ref{eq:stress_function_Q}, we can see that the Hertzian contribution gives rise to nonlinearity as it scales with $Q^\mathrm{3/2}$. Therefore, it is expected that this Hertzian contribution will appear in the third harmonics stress response whereas the capillary force should only appears in the first harmonic stress response ($F_\mathrm{capillary}\sim Q$ to leading order).  
As we only focus on the weak nonlinearity using the third harmonic stress signal $\sigma'_\mathrm{3}$, we use the following fitting equation to analyze our data:
\begin{equation}
\sigma'_\mathrm{3}= 
	\begin{cases}
		-A\left(\gamma_\mathrm{0}+\hat{\gamma}\right)^{3/2}+A \hat{\gamma}^{3/2} & \text{for $\hat{\gamma}> 0$ }\\
		-A\left(\gamma_\mathrm{0}+\hat{\gamma}\right)^{3/2}\mathrm{H}\left(\gamma_\mathrm{0}+\hat{\gamma}\right) & \text{for $\hat{\gamma}\leq0$}
		\end{cases}		
\label{eq:fitting_Hertzian_capillary}
\end{equation}
The offset $\hat{\gamma}$ describes the required applied strain needed to separate the preloaded particles ($\hat{\gamma}>0$) or to bring the particles into contact ($\hat{\gamma}<0$). The parameter $A$ provides the magnitude of the repulsive Hertzian force and $\mathrm{H}\left(\gamma_\mathrm{0}+\hat{\gamma}\right)$ is the Heaviside step function that ensures that the third harmonic elastic stress only arises once the particles are in contact. For $\hat{\gamma}>0$, the $A \hat{\gamma}^{3/2}$ term is added to ensure that $\sigma'_\mathrm{3}=0$ at zero imposed strain ($\gamma_\mathrm{0}=0$); $\sigma'_\mathrm{3}$ is also zero at $\gamma_\mathrm{0}=0$ for $\hat{\gamma}\leq0$ due to the Heaviside function.

It should be noted that while equation~\ref{eq:fitting_Hertzian_capillary} has no explicit length scale, its derivation relied on the assumption that both the repulsive Hertzian contact and capillary attraction occurred between particles and not asperities. A Hertzian contact, regardless of the curvature length scale used (asperity or particle), results in a noninteger force displacement response. Either length scale could be used to rationalize the quantitative observations of the power law scaling, but we chose the radius since the translation from microscopic force to macroscopic stress in the Kramers expression should be linked to the arrangement of the particle network and the radius is more closely linked to the shape of the capillary bridges. 

%%%%  Section Results and discussion %%%%%
\section{Results and discussion} \label{results_discussions}

\subsection{Pendular state}

\begin{figure*}[ht]%
    \centering
    \includegraphics[width=\textwidth]{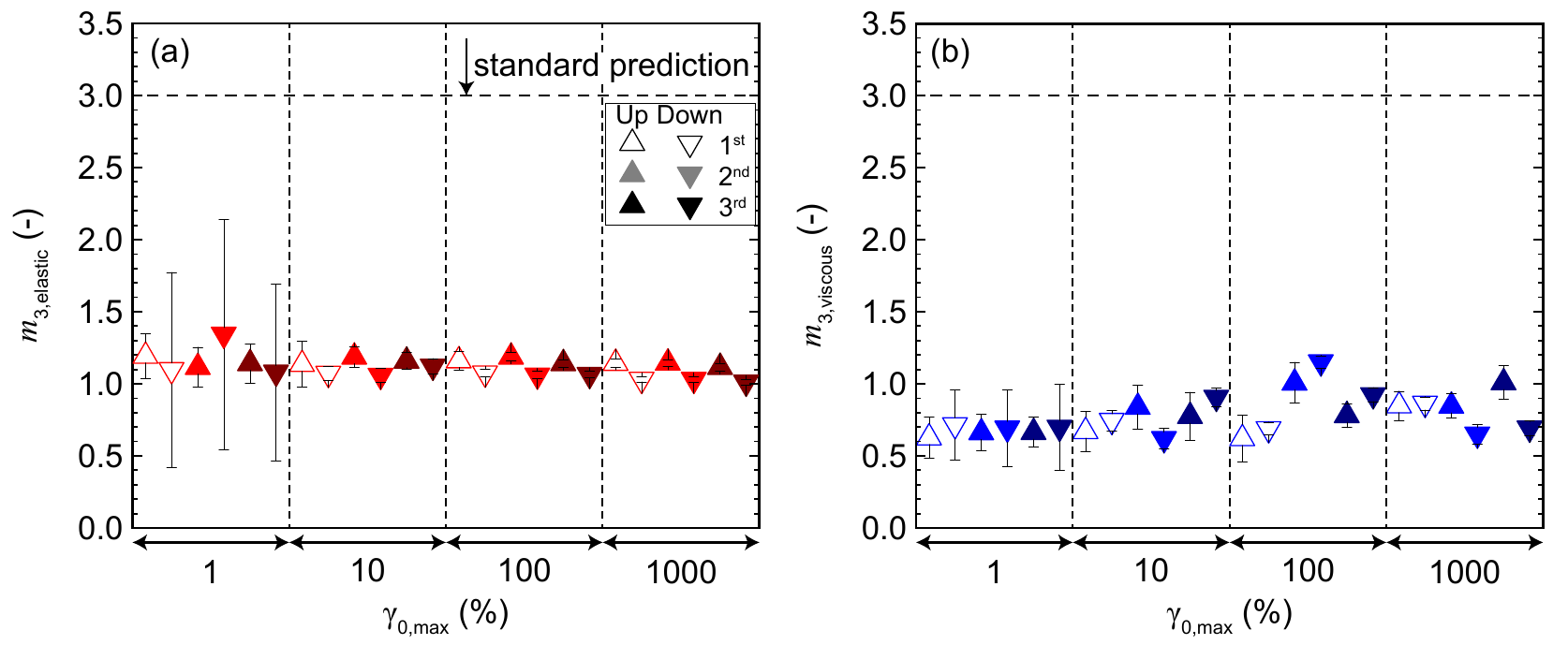}
\caption[noninteger power law scaling for the pendular state $\phi_\mathrm{solid}=0.25$ and $\phi_\mathrm{2nd}/\phi_\mathrm{solid}=0.05$.]{Example of noninteger and distinctive power law scaling from the (a) elastic and (b) viscous third harmonic stresses for a capillary suspension in the pendular state (NP3-silicone~oil-glycerol) with $\phi_\mathrm{solid}=0.25$ and $\phi_\mathrm{2nd}/\phi_\mathrm{solid}=0.05$. The exponent $m_\mathrm{3}$ is the leading order of the nonlinearities, determined from the fitting function of (a) $\sigma'_\mathrm{3}=-[e_\mathrm{3}]\gamma^{m_\mathrm{3,elastic}}$  and (b) $\sigma''_\mathrm{3}=\omega[v_\mathrm{3}]\gamma^{m_\mathrm{3,viscous}}$  for various maximum amplitudes ($\gamma_\mathrm{0}=$1, 10, 100, and 1000$\%$). None of the values matched the standard prediction of the cubic leading order of nonlinearities ($m_\mathrm{3}=3$).  The results for the other compositions are shown in Figure S4.} %\ref{supp-fig:m3_pendular}.}%
\label{fig:m3_pendular-ratio005}%
\end{figure*}

The power law scaling of third harmonic elastic ($m_\mathrm{3,elastic}$) and viscous stresses ($m_\mathrm{3,viscous}$) are determined by fitting equations \ref{eq:sigma_prime_3} and \ref{eq:sigma_doubleprime_3}, neglecting the higher order terms so that $\sigma'_\mathrm{3}=-[e_\mathrm{3}]\gamma_\mathrm{0}^{m_\mathrm{3,elastic}}$ and $\sigma''_\mathrm{3}=\omega[v_\mathrm{3}]\gamma_\mathrm{0}^{m_\mathrm{3,viscous}}$. Details of the fitting procedure are given in the study by Natalia {\it et al.} \cite{Natalia.2020} with an example data set shown in Figure \ref{fig:ExampleMAOS}.
These values, shown as a function of strain amplitude for a pendular state suspension with $\phi_\mathrm{2nd}/\phi_\mathrm{solid}=0.05$, are plotted in Figure \ref{fig:m3_pendular-ratio005} for each measurement subset. The values of the third harmonic elastic scaling (Figure \ref{fig:m3_pendular-ratio005}a) show no dependence on  $\gamma_\mathrm{0,max}$ with an average $ m_\mathrm{3,elastic}=1.08\pm0.13$. It appears that the scalings in the forward order of increasing strain are always slightly higher than the values in the reverse order; however, their difference falls within the uncertainty.
None of the experiments show $m_\mathrm{3}=3$, and a similar trend is observed for other $\phi_\mathrm{2nd}/\phi_\mathrm{solid}$ (see SI, Figure S4).%\ref{supp-fig:m3_pendular}). 
For the viscous scaling, $m_\mathrm{3,viscous}$ is constant at $m_\mathrm{3,viscous}=0.686\pm0.56$ until $\gamma_\mathrm{0,max}=10\%$. At $\gamma_\mathrm{0,max}>10\%$, the values increase slightly as does the scatter between the repetitions. 

\begin{figure*}[htp]%
    \centering
    \includegraphics[width=0.5\textwidth]{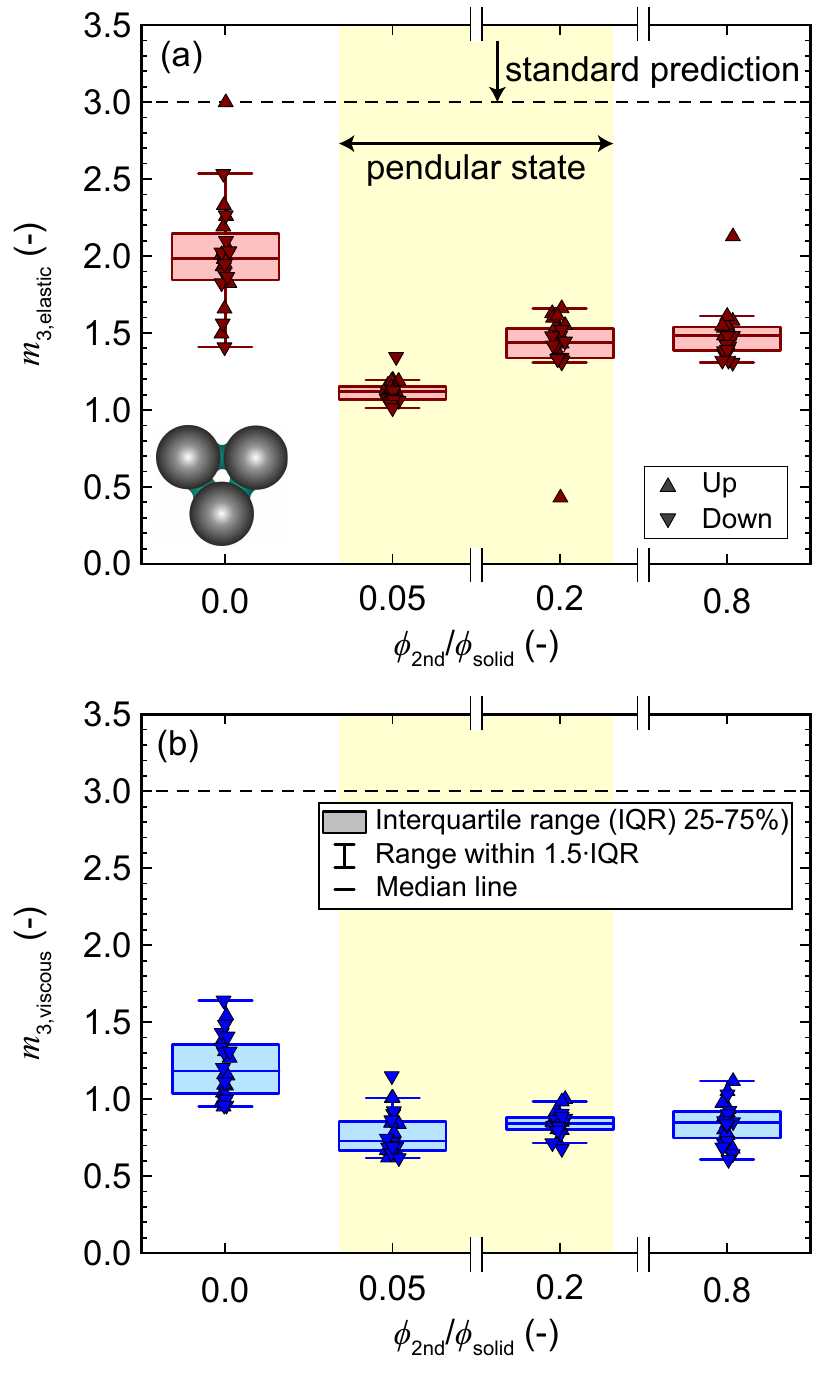}
	\caption[Boxplot representation of the power law scaling from the (a) $\sigma'_\mathrm{3}$ and (b) $\sigma''_\mathrm{3}$ of ternary NP3-silicone oil-glycerin suspensions with different $\phi_\mathrm{2nd}$.]{Boxplot representation of the power law scaling from the (a) elastic and (b) viscous third harmonic stresses of ternary NP3-silicone oil-glycerin suspensions with a constant $\phi_\mathrm{solid}=0.25$ and variation of  $\phi_\mathrm{2nd}/\phi_\mathrm{solid}$. Each boxplot represents fitting values from 24 measurement subsets. The yellow area ($\phi_\mathrm{2nd}/\phi_\mathrm{solid}= 0.05$ and $0.2$) denotes the capillary suspensions in pendular state. The sample with ($\phi_\mathrm{2nd}/\phi_\mathrm{solid}= 0.8$) is in transition between spherical agglomeration and bijel according to the ternary system schematic in Figure \ref{fig:ternary_system}.}%
\label{fig:boxplot_pendular}%
\end{figure*}

To draw a better comparison of the asymptotic nonlinear scaling between different samples, we plot the $m_\mathrm{3}$ distribution of the 24 amplitude sweeps from each sample in a boxplot (Figure \ref{fig:boxplot_pendular}). None of the NP3-silicone oil-glycerol suspensions exhibit the typical cubical scaling ($\sigma_\mathrm{3} \sim \gamma_\mathrm{0}^3$).
The values of $m_\mathrm{3,elastic}$ from the normal suspension ($\phi_\mathrm{2nd}/\phi_\mathrm{solid}=0.0$) have a wide distribution between 1.4 and 3; however, the value 3 is an outlier that is obtained from the first ramp up after the measurement was started from the quiescent state (see SI, Figure S4a). %\ref{supp-fig:m3_pendular-ratio00}). 
Large variance of the raw data from triplicate measurements causes the large uncertainty in the fitting result, and, therefore, this specific data point will not be discussed further.

With the addition of the secondary fluid ($\phi_\mathrm{2nd}/\phi_\mathrm{solid}= 0.05$), the $m_\mathrm{3,elastic}$ values decrease to a minimum before increasing again with higher concentrations of secondary fluid (see Figure \ref{fig:boxplot_pendular}). The values of $m_\mathrm{3,elastic}$ from capillary suspensions at $\phi_\mathrm{2nd}/\phi_\mathrm{solid}= 0.2$ are similar to those with $\phi_\mathrm{2nd}/\phi_\mathrm{solid}= 0.8$, which is expected to be in the transition between the spherical agglomeration and the bicontinuous gel. Like its elastic counterpart, $m_\mathrm{3,viscous}$ is greatest for $\phi_\mathrm{2nd}/\phi_\mathrm{solid}= 0.0$ and decreases at $\phi_\mathrm{2nd}/\phi_\mathrm{solid}= 0.05$. At higher $\phi_\mathrm{2nd}/\phi_\mathrm{solid}$, the change in $m_\mathrm{3,viscous}$ is constant within the interquartile range (IQR). The $m_\mathrm{3,viscous}$ values are smaller than their corresponding $m_\mathrm{3,elastic}$ values. The majority of $m_\mathrm{3,viscous}$ values from the ternary solid-liquid-liquid systems are less than unity (see Figure \ref{fig:boxplot_pendular}b).

\begin{figure}[ht]%
	\centering
	\includegraphics[width=0.5\textwidth]{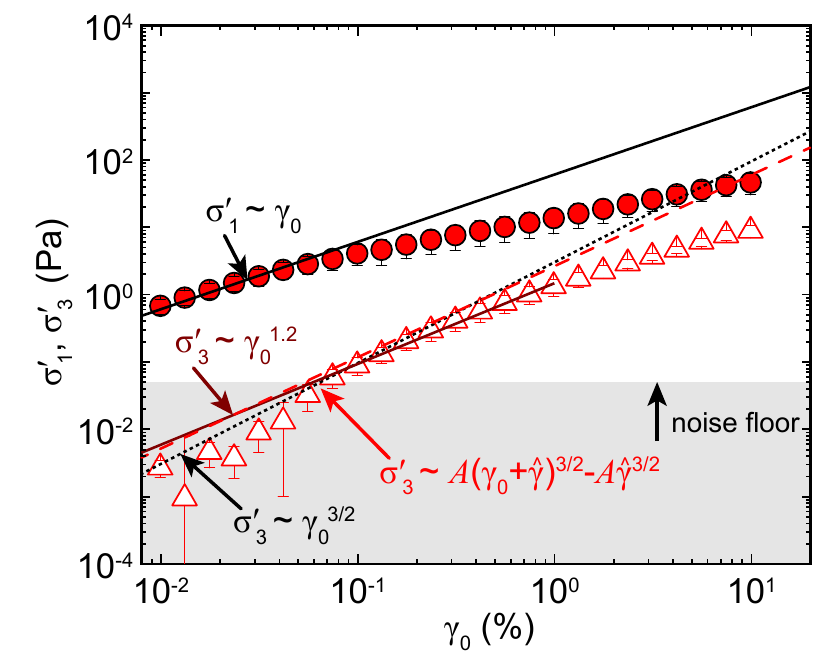}%
	\caption[Fitting comparison for $\sigma'_\mathrm{3}$ in pendular state to explain the noninteger scaling phenomenon.]{Comparison between three fitting functions for third harmonic elastic stresses in the pendular state (NP3-silicone~oil-glycerol). The solid red line denotes the computed power law fitting $\sigma'_3\sim \gamma_\mathrm{0}^{1.2}$. The black dotted line represents the Hertzian repulsive contact force without preloading and the red dashed line represents the fitting equation \ref{eq:fitting_Hertzian_capillary} with preload. 
These data belong to the pendular state sample with $\phi_\mathrm{solid}=0.25$ and $\phi_\mathrm{2nd}/\phi_\mathrm{solid}=0.05$ for the $2^\mathrm{nd}$ up measurement subset with $\gamma_\mathrm{0,max}= 10\%$ at $\omega= 0.628$ rad/s. Open symbols denote negative values and filled symbols positive values. The 0.05 Pa noise floor was determined from the power spectrum (Figure S1b).}% \ref{supp-fig:power_spectrum_stress-pendular005}).}%
	\label{fig:pendular005_09_fitcomparison}%
\end{figure}

A typical elastic data set, a second upward sweep to $\gamma_\mathrm{0,max}=10\%$ for $\phi_\mathrm{2nd}/\phi_\mathrm{solid}= 0.05$, is shown in Figure \ref{fig:pendular005_09_fitcomparison}. The corresponding viscous data  set can be found in SI, Figure S2. %\ref{supp-fig:sigmaDoublePrime1_3_pendular005_7}. 
The first harmonic $\sigma'_\mathrm{1}$ is fit by $\sigma'_\mathrm{1}(\omega,\gamma_\mathrm{0})=G'_\mathrm{LVE}(\omega) \gamma_\mathrm{0}$ in the linear regime (Figure S3). %\ref{supp-fig:G1_LVE_pendular}). 
In the MAOS regime, the third harmonic shows the computed fit 
$\sigma'_\mathrm{3}=-[e_\mathrm{3}]\gamma_\mathrm{0}^{m_\mathrm{3,elastic}}$ (Equation \ref{eq:sigma_prime_3})
with $m_\mathrm{3,elastic}=1.2$. We also show $\sigma'_\mathrm{3} \sim \gamma_\mathrm{0}^{3/2}$, the fit predicted by Equation \ref{eq: Hertzian_force} without preloading ($\hat{\gamma}=0$). For the rigid, but deformable particles, the capillary force causes the particles to be in contact under a preloaded condition, giving rise to the nonlinear elastic response. 

The preloading of the contact by bridges, which occurs during the sample preparation, is affected by the force from the neighboring particles, e.g. the network structure and distribution of bridge and particle sizes. Using the fitting equation \ref{eq:fitting_Hertzian_capillary}, we obtain a value of $\hat{\gamma}=0.03\%$ for the sample shown in Figure \ref{fig:pendular005_09_fitcomparison}. If we ignore the influence of the network structure and bridge size variations, we can calculate the order of magnitude of the particle deformation $\delta$ due to the attractive capillary force using Equation \ref{eq:Hertzian_capillary_force} to check the validity of the fitting result shown in Figure \ref{fig:pendular005_09_fitcomparison}. For the approximation of $\delta$, we use the first term of capillary force only ($F_\mathrm{capillary}=2\pi\Gamma R$), a typical value of Young's modulus of silica glass spheres with $E\sim 80~\mathrm{GPa}$, and the Poisson ratio of $\nu= 0.5$, which results in $\delta\approx 4\cdot 10^{-10}~\mathrm{m}$ or $\delta/R\approx 0.04\%$. 

To check if the standard cubic scaling can appear under the assumption of precompression with the adhesive Hertzian contact, we calculated the Taylor expansion of the fitting equation \ref{eq:fitting_Hertzian_capillary},
\begin{equation}
\sigma'_\mathrm{3}(\gamma_\mathrm{0})\approx -\frac{3}{2}A\hat{\gamma}^{1/2}\gamma_\mathrm{0}-\frac{3A{\gamma_\mathrm{0}}^2}{8\hat{\gamma}^{1/2}}+\frac{A{\gamma_\mathrm{0}}^3}{16\hat{\gamma}^{3/2}}-\frac{3A{\gamma_\mathrm{0}}^4}{128\hat{\gamma}^{5/2}}+\frac{3A{\gamma_\mathrm{0}}^5}{8\hat{\gamma}^{7/2}}+ \mathcal{O}({\gamma_\mathrm{0}}^{6})
	\label{eq:taylor_exp_jkr_model}
\end{equation}
A Taylor series for Equation \ref{eq:fitting_Hertzian_capillary} is only possible for finite precompression ($\hat{\gamma}\neq0$), and it is restricted to very small $\gamma_0$.
Equation \ref{eq:taylor_exp_jkr_model} shows that at certain $\hat{\gamma}$, all higher order terms become important at the same time and, therefore, the full form Equation \ref{eq:fitting_Hertzian_capillary} has to be used. Using the first two higher orders of the Taylor expansion, we can determine the limit for $\gamma_\mathrm{0}$ to truncate the expansion at the first nonlinearity,
\begin{equation*}
 \frac{A{\gamma_\mathrm{0}}^3}{16\hat{\gamma}^{3/2}} \ll \frac{3A{\gamma_\mathrm{0}}^2}{8\hat{\gamma}^{1/2}}
\label{eq:first_two_orders}
\end{equation*}
\begin{equation}
 \gamma_\mathrm{0} \ll 6\hat{\gamma}
\end{equation}
A similar result appears when comparing any adjacent higher order terms, but with a different numerical prefactor. Thus, in order to see the integer scaling, the strain amplitude input $\gamma_\mathrm{0}$ has to be extremely small compared to $\hat{\gamma}$. Taking the fitting result of $\hat{\gamma}=0.03\%$ from Figure \ref{fig:pendular005_09_fitcomparison}, the region with a cubical scaling ($\gamma_\mathrm{0}\ll0.03\%$) was not probed, as the smallest strain used in this measurement is $\gamma_\mathrm{0}=0.01\%$. Furthermore, current limitations in instrument design make such measurements unreliable due to the precision of the displacement and torque sensors.
Using much softer particles, which have stronger precompression, might help make integer scaling observable as this could shift the limit on imposed strain to a value above the displacement limit. For the limit of no precompression, using dry granular materials with Hertzian contact should be considered in the future work.

While the power law scaling with $m_\mathrm{3,elastic}=1.2$ appears to fit the data above the noise floor better than the Hertzian contact with precompression (Eq.~\ref{eq:fitting_Hertzian_capillary}) in Figure~\ref{fig:pendular005_09_fitcomparison}, the fit has two clear advantages. First, the fit has a clear underlying physical basis that can be used to make predictions about the system. Second, the present model implies the existence of both an apparent cubic scaling in the present materials at low deformations ($ \gamma_\mathrm{0} \ll 6\hat{\gamma}$) with a transition towards $m_\mathrm{3,elastic} = 1.5$ at higher deformations. While we typically define the asymptotically nonlinear MAOS regime as the region where the shear stress response becomes nonlinear with the appearance of a third harmonic while all higher order harmonics are negligibly small~\cite{Ewoldt.2013}, the inherent nonlinearity of the Hertzian contact with precompression would imply that the third (and even higher harmonics) arise at very small strains, well beyond what we can measure using even the most sensitive rheometers at present. That said, the fifth harmonic is always lower than the third harmonic, even at strain amplitudes above the MAOS region. At strain amplitude of 10\%, the magnitude of the fifth harmonic is approximately a half order of magnitude lower than the third harmonic and is even lower than the apparent second harmonic (the second harmonic is often considered a proxy for noise). The analyzed MAOS regime is typically at lower strains, with even lower relative fifth harmonics, e.g.~for the data shown in Figure~\ref{fig:pendular005_09_fitcomparison} ($\phi_\mathrm{solid}=0.25$, $\phi_\mathrm{2nd}/\phi_\mathrm{solid}=0.05$), we found that the MAOS region ended at approximately $\gamma_\mathrm{0}=1\%$.

\begin{figure*}[htp!]%
	\centering
	\includegraphics[width=0.5\textwidth]{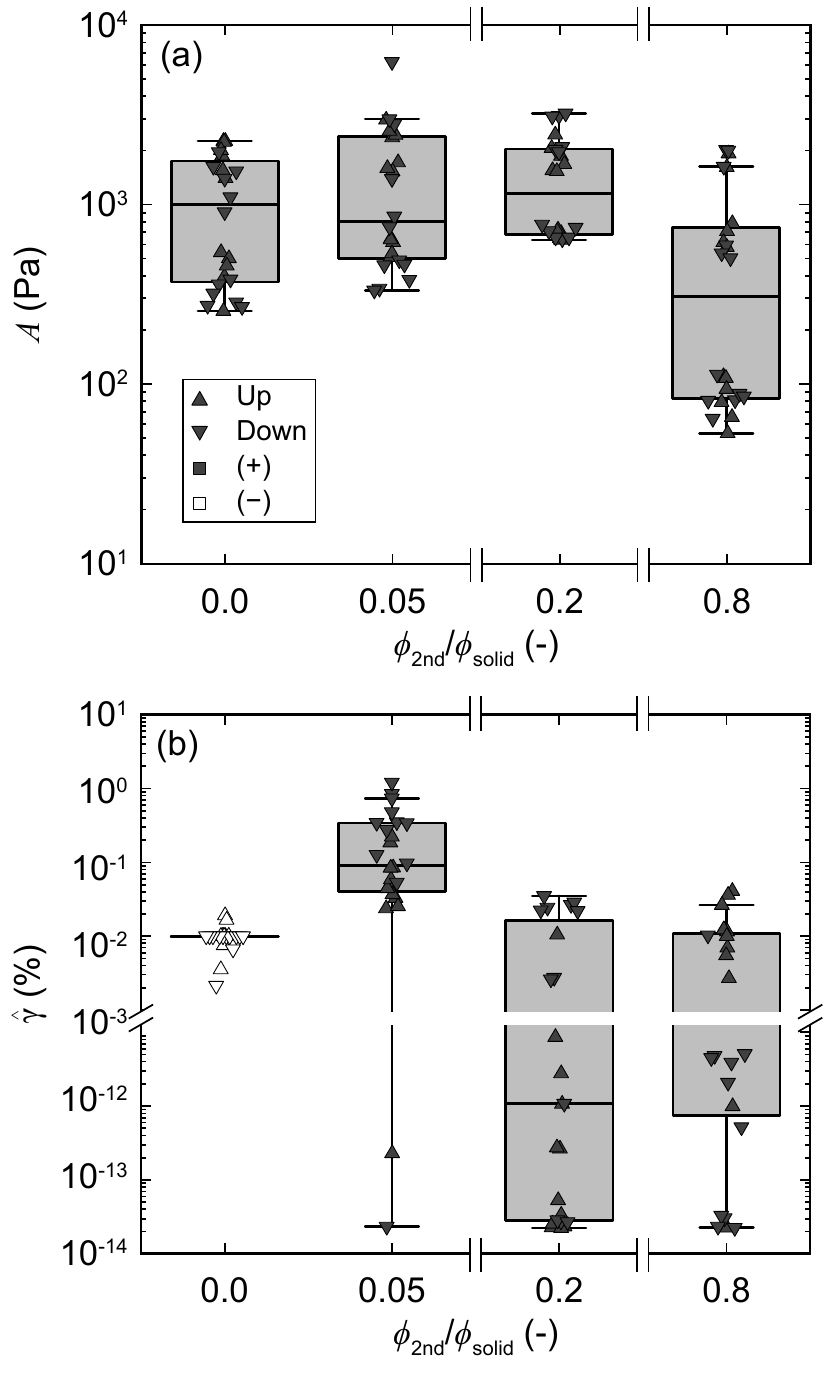}%
	\caption[Fitting results from the Equation \ref{eq:fitting_Hertzian_capillary} for pendular state samples with different concentrations of secondary fluid.]{Fitting results (a) $A$ and (b) $\hat{\gamma}$ from the Equation \ref{eq:fitting_Hertzian_capillary} for samples with different concentration of secondary fluid. Negative values of $\hat{\gamma}$ are denoted using open symbols.}%
\label{fig:JKR_pendular_fitting_paramater_boxplot}%
\end{figure*}

The fitting parameters $A$ and $\hat{\gamma}$ are plotted as a function of the secondary fluid concentration ($\phi_\mathrm{2nd}/\phi_\mathrm{solid}$) in Figure \ref{fig:JKR_pendular_fitting_paramater_boxplot}. Their dependence on $\gamma_\mathrm{0,max}$ is shown in SI, Figure S5 and tabulated in Table S1. % \ref{supp-fig:JKR_pendular_fitting_paramater}.
The numerical value of $A$ depends on the units of strain used in equation~\ref{eq:fitting_Hertzian_capillary}. A fitting parameter comparison using strain amplitude input in percentage unit ($\gamma[\%]$) and unitless ($\gamma[-]$) shows that the value $A$ shifts by a factor of $(10^2)^{1.5}=10^3$ and the value $\hat{\gamma}$ by a factor of $10^2$, as expected from units conversion. 
Therefore, the reported $A$ values in this paper, associated with $\%$-unit in the strain amplitude, would need to be increased by a factor of 1000 if standard strain units are used for $\gamma$ instead.
Parameter $A$ shows no clear trend between normal suspension and capillary suspensions ($\phi_\mathrm{2nd}/\phi_\mathrm{solid}=0.05$ -- $0.2$), indicating similar repulsive responses between these three sample compositions. The bicontinuous sample with $\phi_\mathrm{2nd}/\phi_\mathrm{solid}=0.8$, however, has a broader distribution in the parameter $A$ and some of its values are lower compared to other sample compositions (see Figure \ref{fig:JKR_pendular_fitting_paramater_boxplot}a). From Equations \ref{eq:Hertzian_capillary_force} and \ref{eq:fitting_Hertzian_capillary}, the reduced Young's modulus $E^*$ governs the fitting parameter $A$. With the increasing concentration of secondary fluid, larger aggregates form and the distance between particles can increase even though they are still ``trapped'' in the cluster. 
The lower effective $E^*$, therefore,  might be an indication of cluster-cluster contacts instead of the more simplistic particle-particle contacts. This results in a drop in $A$ since the clusters are effectively softer than the single particles.

Without the addition of the secondary fluid ($\phi_\mathrm{2nd}/\phi_\mathrm{solid}=0.0$), there is no deformation of the particle surface at rest, as can be seen in the negative value of $\hat{\gamma}$ (open symbols on Figure \ref{fig:JKR_pendular_fitting_paramater_boxplot}b). The magnitude of negative $\hat{\gamma}$ depicts the strain required to induce contact between particles. The vdW force between particles is typically 1--3 orders of magnitude weaker than the capillary force \cite{Koos.2014}; therefore, it is insufficient to deform the surface of silica spheres. With the addition of the secondary fluid ($\phi_\mathrm{2nd}/\phi_\mathrm{solid}=0.05$), the $\hat{\gamma}$ values are positive and increase through maximum before decreasing again to infinitesimally small values at higher concentrations of secondary fluid ($\phi_\mathrm{2nd}/\phi_\mathrm{solid}=0.2$ and 0.8). These infinitesimally small values $\hat{\gamma} \approx 10^{-11}$ -- $10^{-14}$ are likely computational artifacts and thus the points should be treated as if they are indistinguishable from zero.  This decline of the $\hat{\gamma}$ values indicates a transition from a large number of binary bridges to fewer funicular clusters ($\phi_\mathrm{2nd}/\phi_\mathrm{solid}=0.2$)  than a bicontinuous gel ($\phi_\mathrm{2nd}/\phi_\mathrm{solid}=0.8$) where the bridges coalesce into larger clusters with particles immersed in the secondary fluid, resulting in the particles being just in contact $\hat{\gamma}\approx0$ for the majority of the of fitting results. 

While the fitting parameter $\hat{\gamma}$ shows a trend between capillary suspension samples with different $\phi_\mathrm{2nd}/\phi_\mathrm{solid}$, the points within each population do not seem to be constant. Samples with $\phi_\mathrm{2nd}/\phi_\mathrm{solid}=0.05$ and 0.2 seem to show higher preloading in the ramp down measurement subsets (shown as downward triangles). However, the opposite trend holds for the sample with $\phi_\mathrm{2nd}/\phi_\mathrm{solid}=0.8$. While more accurate measurements may confirm this trend, we assume that the limited data points above the noise floor cause these alternating values, as the fitting function is very sensitive to any small deviation of the data points.

\begin{figure}[ht]%
		\centering
		\includegraphics[width=0.5\textwidth]{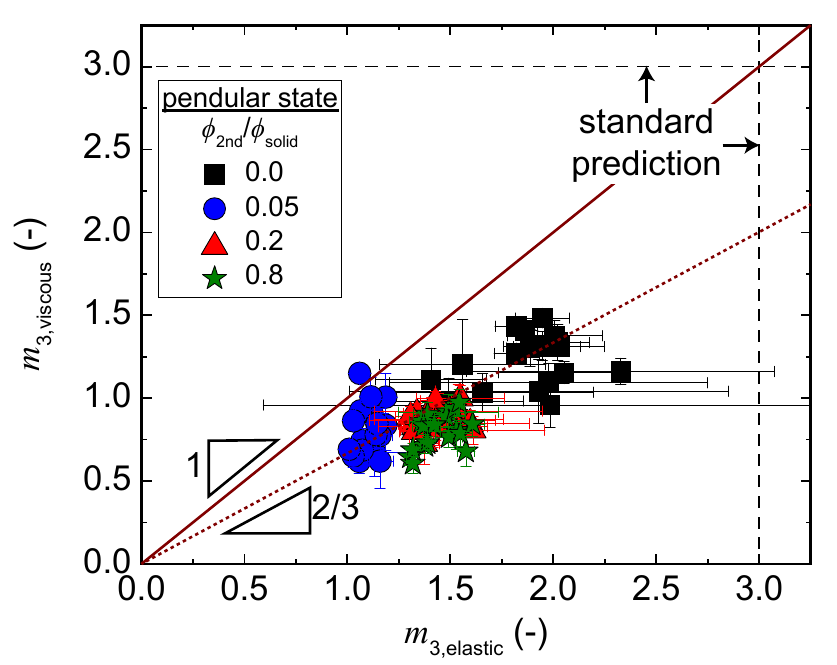}%
	\caption[Relationship between $m_\mathrm{3,viscous}$ and $m_\mathrm{3,elastic}$ in pendular state.]{For all pendular state samples (NP3-silicone~oil-glycerol), $m_\mathrm{3,viscous}$ is smaller than the $m_\mathrm{3,elastic}$ which might be related to the adhesion-controlled friction. The adhesion between particles comes from the attractive forces.}%
\label{fig:elasticVSviscous_pendular}%
\end{figure}

As mentioned in the Introduction, the goal of our work is to understand the physical origin of the noninteger power law exponents $m_\mathrm{3}$ for the third harmonic elastic as well as viscous stresses, and explain why $m_\mathrm{3,viscous} < m_\mathrm{3,elastic}$ for our reported samples (see Figure \ref{fig:elasticVSviscous_pendular} and Table S2). %\ref{supp-table:ratio_viscous_to_elastic_scaling}). 
In the following paragraphs, we propose how $m_\mathrm{3,viscous}$ can be linked to the adhesion-controlled friction.

Under ideal conditions, both the stretching of the bridge and the Hertzian compression are purely elastic. Under real conditions, however, a contact angle hysteresis can cause a difference between the advancing and receding contacts (as well as the displacement speed dependence). Furthermore, contact angle pinning and the influence of the network structure  can cause a hysteresis in the load-contact radius relationship. This hysteresis causes an energy dissipation, which predominantly occurs in the form of friction. During frictional sliding, the particles are not fully separated, due to the attractive force, in contrast to nonadhesive contact. Thus, the energy dissipation at rest is higher for an adhesive contact compared to a nonadhesive contact. Yet, under deformation, the presence of adhesive contact decreases the energy dissipation rate \cite{Israelachvili.2011}. This confirms the results from Kovalcinova {\it et al.} where wet granular matter exhibits lower energy dissipation rate than dry granular matter during shear\cite{Kovalcinova.2018}. 

The adhesion-controlled friction, which has a strong dependence on the contact area, dominates the friction at low applied load. Consequently, the linear relationship between the load and friction (Amonton's law) in load-controlled friction is no longer valid \cite{Gao.2004, Xu.2007}. Based on their experimental results, Riedo {\it et al.} reported a 2/3 power law dependence of the friction force on the normal load that originates from the load dependence of the contact area \cite{Riedo.2004}. In our case, the attractive capillary force is the origin of the normal load. 
Since both $m_\mathrm{3,elastic}$ and $m_\mathrm{3,viscous}$ are related to the capillary force through $\hat{\gamma}$ in the elastic scaling and the adhesive-controlled friction in the viscous scaling, we therefore expect that $F_\mathrm{friction}\sim F_\mathrm{capillary}^{2/3}$ or, for the sample without added secondary fluid, $F_\mathrm{friction}\sim F_\mathrm{vdW}^{2/3}$. We can link the elastic scaling to the strength of the normal load and the viscous scaling to the friction, thus, $m_\mathrm{3,viscous}\approx \frac{2}{3}m_\mathrm{3,elastic}$. This relationship is shown in Figure \ref{fig:elasticVSviscous_pendular}. This relationship between the two scalings would explain the observation that $m_\mathrm{3,viscous}$ is always smaller than $m_\mathrm{3,elastic}$. For clarity, scalings for $\gamma_\mathrm{0,max}=1\%$ are not shown due to the large uncertainty of the fitting values caused by the limited raw data above the torque limit in this regime. While the scaling for $m_\mathrm{3,viscous}\approx \frac{2}{3}m_\mathrm{3,elastic}$ holds between the different $\phi_\mathrm{2nd}/\phi_\mathrm{solid}$ samples, the points within each population do not seem to follow this scaling. 

%%% Capillary state
\subsection{Capillary state}
\begin{figure*}[htp]%
		\centering
		\includegraphics[width=0.5\textwidth]{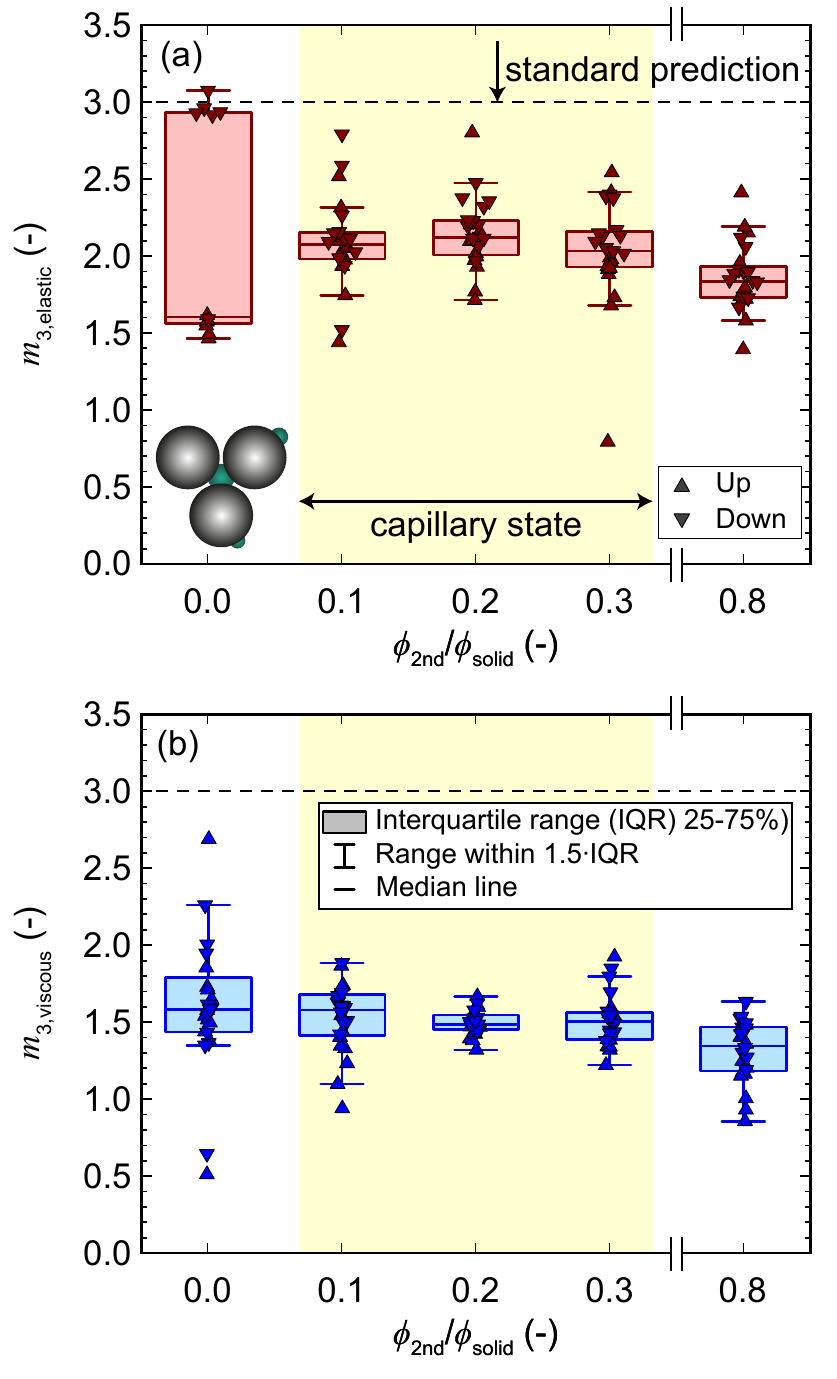}%
	\caption[Boxplot representation of the power law scaling from the (a) $\sigma'_\mathrm{3}$ and (b) $\sigma''_\mathrm{3}$ of ternary PMMA-glycerol-paraffin oil suspensions with different $\phi_\mathrm{2nd}$.]{Boxplot representation of power law scaling from the (a) elastic and (b) viscous third harmonic stresses of ternary PMMA-glycerin-paraffin oil suspensions  with a constant $\phi_\mathrm{solid}=0.25$ and variation of  $\phi_\mathrm{2nd}/\phi_\mathrm{solid}$. The yellow area ($\phi_\mathrm{2nd}/\phi_\mathrm{solid}= 0.1$ -- $0.3$) denotes the capillary suspensions in the capillary state and sample with ($\phi_\mathrm{2nd}/\phi_\mathrm{solid}= 0.8$) refers to a Pickering emulsion-like sample according to the ternary system schematic in Figure \ref{fig:ternary_system}.}%
\label{fig:boxplot_capillary}%
\end{figure*}

In the capillary state, the liquid bridges have convex menisci and the particles form a cluster around the secondary fluid droplets. Samples with PMMA in glycerol with added paraffin oil were used as model systems in the capillary state. As before, we plot the elastic and viscous third harmonic scalings ($m_\mathrm{3,elastic}$ and $m_\mathrm{3,viscous}$) of the capillary state samples as a function of secondary fluid concentration in a boxplot representation (Figure \ref{fig:boxplot_capillary}). This model system shows higher scaling values and a broader distribution for each sample composition compared to the pendular state suspensions. The shear moduli in the linear regime are plotted in the SI Figure S8 %\ref{supp-fig:G1_LVE_capillary} 
along with the details of elastic and viscous scaling as a function of maximum strain amplitude ($\gamma_\mathrm{0,max}$) in Figure S9. % \ref{supp-fig:m3_capillary}.

The normal suspension ($\phi_\mathrm{2nd}/\phi_\mathrm{solid}=0.0$) exhibits a clear bimodal scaling distribution, with one cluster of $m_\mathrm{3,elastic}$ around 3 and the other around 1.5. This is the only sample in this paper that shows the typical cubical scaling. Furthermore, this is the only sample composition in this paper that shows a viscous dominated behavior ($G''_1>G'_1$) for all ramping up and down amplitude sweep measurement subsets. Due to the lower viscosity of this sample, the measurement window at low strain amplitude is limited since $G''_1(\gamma_\mathrm{0}\leq 0.1\%)$ and $G'_1(\gamma_\mathrm{0}\leq 10\%)$ fall below the torque limit.  
Hence, only data sets with $\gamma_\mathrm{0,max}\geq 100\%$ are considered for the calculation of $m_\mathrm{3,elastic}$ and data sets with $\gamma_\mathrm{0,max}\geq 10\%$ for the calculation of $m_\mathrm{3,viscous}$. This resulted in fewer data points for the determination of the elastic and viscous scaling in comparison to other sample compositions. For the details of the measurement limit for this sample, see SI, Figure S7. %\ref{supp-fig:G1_capillary_00}.
Interestingly, the $m_\mathrm{3,elastic}$ scaling alternates between $m_\mathrm{3,elastic}\sim 3$ for the ramp down measurement subsets and $m_\mathrm{3,elastic}\sim 1.5$ for the ramp up measurement subsets in Figure \ref{fig:boxplot_capillary}a (see also Figure S9a). %\ref{supp-fig:m3_capillary-ratio00}). 
This trend implies a change in the dynamics and/or structure between the increasing and decreasing amplitude measurements. This alternating scaling, however, is not evident in the viscous scaling measurements (see Figure \ref{fig:boxplot_capillary}b and Figure S9a). %\ref{supp-fig:m3_capillary-ratio00}). 

The outliers of the elastic scaling distribution for other samples belong to the first six ramping up and down measurement subsets ($\gamma_\mathrm{0,max} = 1\%$), due to the limited data points above the $\sigma_\mathrm{min}$ that leads to a larger uncertainty in the fitting (see SI Figure S9). % \ref{supp-fig:m3_capillary}).
The elastic scaling of capillary suspensions in the capillary state ($\phi_\mathrm{2nd}/\phi_\mathrm{solid}=0.1-0.3$) is nearly independent of the secondary fluid concentration. The interquartile range (IQR) of the elastic scaling distribution of these compositions is found between 1.9 and 2.3. With an increasing amount of secondary fluid ($\phi_\mathrm{2nd}/\phi_\mathrm{solid}=0.8$), denoting the transition to a Pickering emulsion-like state, the elastic scaling decreases and the IQR shifts to values between 1.7 and 1.9. Interestingly, no cubical viscous scaling is observed, not even for the normal suspension that periodically shows elastic cubical scaling (IQR of $m_\mathrm{3,viscous}=1.4-1.8)$. With the addition of the secondary fluid, $m_\mathrm{3,viscous}$ decreases and remains constant for $\phi_\mathrm{2nd}/\phi_\mathrm{solid}=0.1-0.3$ with IQR of $m_\mathrm{3,viscous}=1.4-1.7$. The viscous scaling reaches the lowest value for $\phi_\mathrm{2nd}/\phi_\mathrm{solid}=0.8$ (IQR between 1.2-1.5).

Similar to the pendular state samples, we use the fitting Equation \ref{eq:fitting_Hertzian_capillary} to understand the cause of the noninteger scaling of the capillary state samples and to determine why the capillary state scaling differs from the pendular state scaling. As has been shown in the theoretical work of Megias-Alguacil and Gauckler, liquid bridges with convex menisci have lower attractive capillary force compared to liquid bridges with concave menisci due to the positive Laplace pressure \cite{MegiasAlguacil.2010}. Therefore, capillary suspensions in the capillary state, with a convex meniscus, are typically reported to have lower yield stress values in comparison to their corresponding pendular state suspensions \cite{Bossler.2016}. Indeed, the yield stress of PMMA-glycerol-paraffin oil based capillary suspensions in the capillary state are reported to be in the same order of the NP3 in silicone oil normal suspension, indicating a weak attractive capillary force \cite{Natalia.2018}. 
In this previous work, no apparent yield stress is reported for the PMMA in glycerol normal suspension, which justifies the assumption that no attractive van der Waals force acts on the particles. 

\begin{figure}[ht]%
\centering
\includegraphics[width=0.5\textwidth]{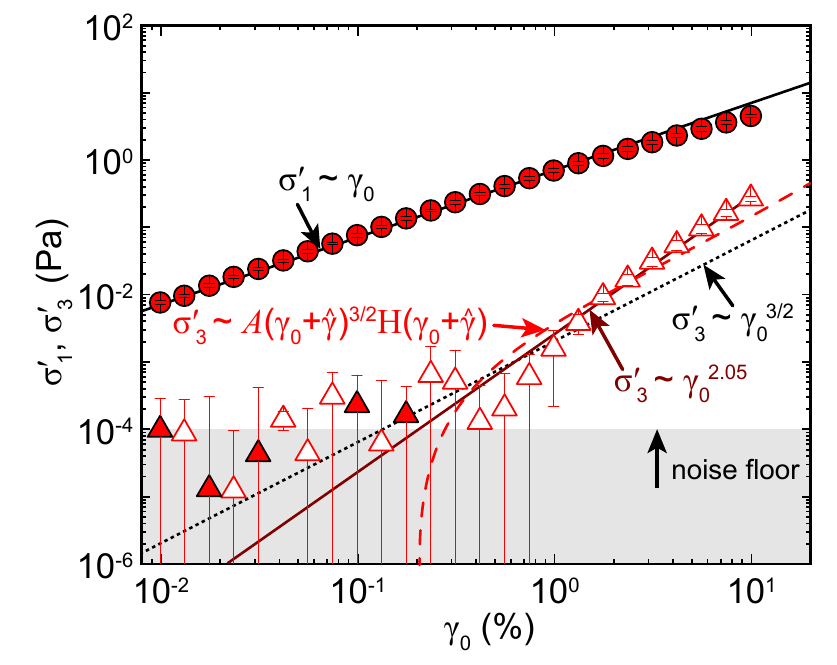}%
\caption[Fitting comparison for $\sigma'_\mathrm{3}$ in capillary state to explain the noninteger scaling phenomenon.]{Comparison between three fitting functions for third harmonic elastic stresses in the capillary state (PMMA-glycerol-paraffin oil).	The solid red line denotes the computed power law fitting $\sigma'_3\sim \gamma_\mathrm{0}^{2.05}$. The black dotted line represents the pure Hertzian repulsive contact force and the red dashed line represents fitting equation $\sigma'_\mathrm{3}=-A\left(\gamma_\mathrm{0}+\hat{\gamma}\right)^{3/2}\mathrm{H}\left(\gamma_\mathrm{0}+\hat{\gamma}\right)$. Open symbols denote negative values and filled symbols positive values. The $10^{-4}$ Pa noise floor was determined from the power spectrum in the study by Natalia {\it et al.} \cite{Natalia.2020}.}%
\label{fig:capillary01_17_fitcomparison}%
\end{figure}

A representative fit for capillary state samples is shown in Figure \ref{fig:capillary01_17_fitcomparison} for the sample with $\phi_\mathrm{2nd}/\phi_\mathrm{solid}=0.1$ at $\gamma_\mathrm{0,max}=10\%$ ($2^\mathrm{nd}$ up measurement). These are the same data shown previously in Figure \ref{fig:ExampleMAOS}a. The data are fit with the function $\sigma'_\mathrm{3}=-A\left(\gamma_\mathrm{0}+\hat{\gamma}\right)^{3/2}\mathrm{H}\left(\gamma_\mathrm{0}+\hat{\gamma}\right)$. As a reminder, $\mathrm{H}\left(\gamma_\mathrm{0}+\hat{\gamma}\right)$ is the Heaviside step function to ensure $\sigma'_\mathrm{3}=0$ before the particles are in contact. For comparison, the Hertzian fit with $\sigma'_\mathrm{3}\sim \gamma_\mathrm{0}^{3/2}$ and the computed single power law of $\sigma'_\mathrm{3} \sim \gamma_\mathrm{0}^{2.05}$ are also shown. Figure \ref{fig:capillary01_17_fitcomparison} also clearly shows the large LVE range where $\sigma'_\mathrm{3}$ only begins to increase above the torque measurement noise floor around $\gamma_\mathrm{0}=1\%$. 

\begin{figure*}[htp]%
	\centering
	\includegraphics[width=0.5\textwidth]{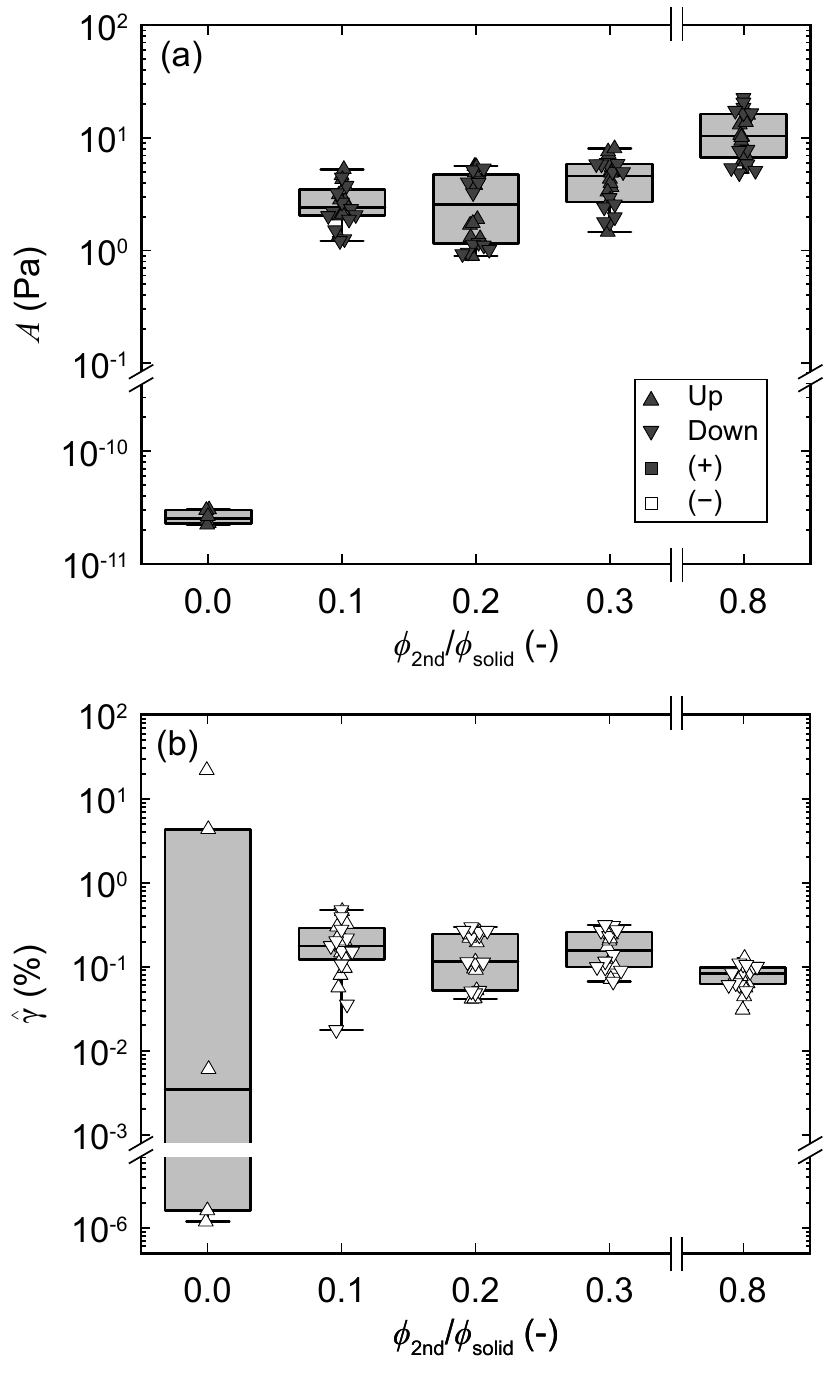}\hfill%
	\caption[Fitting results from the fitting equation \ref{eq:fitting_Hertzian_capillary} for samples with different concentration of secondary fluid in the capillary state.]{Fitting results from the fitting equation $\sigma'_\mathrm{3}=-A\left(\gamma_\mathrm{0}+\hat{\gamma}\right)^{3/2}\mathrm{H}\left(\gamma_\mathrm{0}+\hat{\gamma}\right)$ for samples with different concentrations of secondary fluid in the capillary state. (a) A and (b) $\hat{\gamma}$. Hollow symbols in $\hat{\gamma}$ denote negative values, indicating the strain required to bring the particles into contact. }%
\label{fig:JKR_capillary_fitting_paramater_boxplot}%
\end{figure*}

The ramping down measurement subsets with the elastic scaling of $\mathrm{m_{3,elastic}}\sim 3$ are not used for the Hertzian fitting comparison, as this fitting equation fails to describe the cubical elastic scaling.
The magnitude of the Hertzian-like contact $A$ values for the normal suspension are infinitely small, consistent with the very weak interaction between the particles (Figure \ref{fig:JKR_capillary_fitting_paramater_boxplot}a, Table S3). Due to the low particle concentration ($\phi_\mathrm{solid}=0.25$) and weak interactions, the particles are distributed fairly evenly at low strain amplitudes. With increasing strain and hydrodynamic force, these particles can collide and repel from each other with a Hertzian force, raising the elastic third harmonic scaling to the power 1.5. This idea is confirmed by the large negative magnitude of $\hat{\gamma}$ (Figure \ref{fig:JKR_capillary_fitting_paramater_boxplot}b), indicating that a large strain is needed to bring the particles into contact. However, this hypothesis for the normal suspension should be confirmed with other measurement techniques since we reach the limitation of the method for this specific sample.

With the addition of the secondary fluid, the parameter $A$ shows an increasing trend proportional to the concentration of secondary fluid (Figure \ref{fig:JKR_capillary_fitting_paramater_boxplot}a), whereas the magnitudes of the negative $\hat{\gamma}$ values show a constant trend for $\phi_\mathrm{2nd}/\phi_\mathrm{solid}=0.1-0.3$ and decrease for $\phi_\mathrm{2nd}/\phi_\mathrm{solid}=0.8$ (Figure \ref{fig:JKR_capillary_fitting_paramater_boxplot}b). As a reminder, a negative $\hat{\gamma}$ indicates that the particles are not yet in contact. The decreasing trend of $\hat{\gamma}$ implies that the interparticle distance decreases as more particles aggregate around the larger secondary fluid droplets to minimize energy. 
Confocal microscope images (see SI, Figure S6)  %\ref{supp-fig:capillary_state_confocal}) 
show that the droplet size and number increase with increasing concentration of secondary fluid. A larger droplet can entrap more particles and, therefore, the interparticle distance within the cluster can also decrease. However, it is hard to see this implication in the value of $\hat{\gamma}$.
A smaller interparticle distance indicates that particle-particle contacts instead of ``softer'' cluster-cluster contacts will emerge, leading to a higher reduced Young's modulus $E^*$ governing the parameter $A$. Values of the fitting parameters for each maximum applied strain are shown in Figure S10. % \ref{supp-fig:DMT_capillary_fitting_paramater}.

\begin{figure}[ht]%
\centering
	\includegraphics[width=0.5\textwidth]{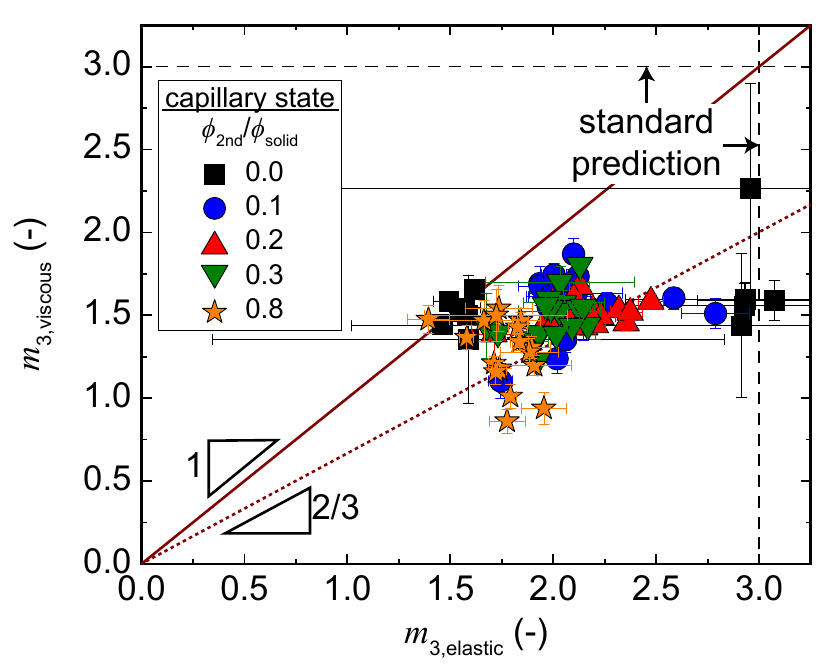}%
\caption[Correlation between $m_\mathrm{3,elastic}$ and $m_\mathrm{3,viscous}$ in the capillary state.]{Correlation between $m_\mathrm{3,elastic}$ and $m_\mathrm{3,viscous}$ in capillary state (PMMA-glycerol-paraffin~oil) samples. Similar to the pendular state samples, $m_\mathrm{3,viscous}$ is smaller than the $m_\mathrm{3,elastic}$ for all sample compositions and this different scaling might be caused by the presence of attractive force and friction.}%
\label{fig:capillary_summary}%
\end{figure}

Like the NP3-silicone oil-glycerol suspensions, the viscous scaling is always lower than the elastic scaling for all samples in the PMMA-glycerol-paraffin oil suspensions (Figure \ref{fig:capillary_summary}). The only exception is for half of the normal suspension samples ($\phi_\mathrm{2nd}=0$), which show the expected $m_\mathrm{3,elastic} \approx 3$. A trend in the viscous scaling $m_\mathrm{3,viscous}$ is difficult to determine, but it appears to be constant with $m_\mathrm{3,elastic}$ and independent from the concentration of secondary fluid. Therefore, we cannot make a correlation for this model system between adhesion and friction, which is expected to be the root of the peculiar noninteger asymptotic nonlinear scaling. The nature of friction in the capillary state remains an unanswered question and needs to be investigated further in the future.

\section{Conclusions} \label{conclusions}

We report an extensive experimental data set of atypical noncubic and noninteger power law scalings for weak nonlinearity over various formulations of capillary suspensions in both the pendular and capillary states. We show that the underlying physics of Hertzian particle contacts is responsible for these as-yet-unexplained noninteger power laws $\sigma_\mathrm{3} \sim \gamma_\mathrm{0}^{m_3}$ with $m_\mathrm{3} \neq 3$ in weakly nonlinear rheology. The model specifically predicts values of the elastic power law exponent $m_{3,elastic}=1.5$, as well as the possibility of a transition to $m_{3,elastic}=3$ at extremely low strain amplitude when the equilibrium particle contact network is preloaded due to the attractive interactions. 
Furthermore, the model provides an interpretation of the ratio $m_{3,viscous}/m_{3,elastic}$ in terms of the type of friction involved, with adhesive-controlled friction found in the present systems. While the model does not predict the MAOS front factor quantitatively, it does rationalize the observed values from fitting the data. 
%Further, we observe that the viscous scaling $ m_\mathrm{3,viscous}$ is always lower than the elastic scaling $ m_\mathrm{3,elastic}$. 

%weakly nonlinear oscillatory shear often correlates with the microstructure and we found that the addition of a physical model for these scaling can give us clear insights into the particle interactions in this material.

%We show that the elastic scaling likely originates from a combination of the Hertzian contact force and the capillary force, where a Hertzian compression between particles occurs because of the attractive capillary force. The coupling between particle contact and friction is postulated to be the origin of the noninteger viscous scaling. Fitting parameters for the strength of the relevant forces (Hertzian, capillary, and friction) provide information about the microstructure for each sample composition.

More specifically, the suspensions in the pendular state exhibit $ m_\mathrm{3,elastic}\leq1.5$. The attractive capillary force causes the deformable particles to be in contact under a preloaded condition. This is clearly evident from the initial deformation $\hat{\gamma}>0$ for $\phi_\mathrm{2nd}/\phi_\mathrm{solid}=0.05$. With increasing secondary fluid concentration, some toroidal bridges coalescence, which results in a weaker capillary force. Thus, the particle-particle contact radius decreases to an infinitesimally small value and the particles are just in contact. The majority of viscous scalings in the pendular state sample are less than unity, which we attribute to the friction at low load. The friction is proportional to the contact area, which has a 2/3 power law dependence on the capillary force as the source of normal load. This is consistent with our observation of $m_\mathrm{3,viscous}\approx\frac{2}{3} m_\mathrm{3,elastic}$.

In contrast, suspensions in the capillary state exhibit $ m_\mathrm{3,elastic}>1.5$. In the capillary state, the liquid bridges have convex menisci and the particles aggregate around the secondary fluid droplets to minimize energy. Although the particles are pinned in the cluster, they are not yet in contact, as reflected in the negative values of prestrain $\hat{\gamma}$. 
This delayed contact, however, provides difficulty linking the origin of the friction in the capillary state samples to the particle contacts, especially since $m_\mathrm{3,viscous}$ is nearly independent from secondary fluid concentration. Friction in noncolloidal suspensions remains a well-posed experimental challenge that now deserves more attention in future work.

As mentioned in our previous work, noninteger scaling for weak nonlinearity has been reported in other material systems, e.g. percolated nanocomposite \cite{Kadar.2017}, concentrated noncolloidal PMMA suspensions \cite{Nam.2011}, Carbopol microgel particle suspension \cite{Blackwell.2016}, and long branched high density polyethylene (HDPE) polymer melt \cite{Hyun.2007}. With the one exception of long branched HDPE, all other materials, including capillary suspensions, are kinetically trapped systems where contact between particles would occur. Therefore, our proposed adhesion-friction theory might also serve well to describe their observed atypical, noninteger MAOS scaling. For example, the results of Nam {\it et al.} \cite{Nam.2011} showed a scaling of 1.5 in the combined measure of $I_{3/1}$. Their concentrated PMMA suspension is consistent with the Hertzian model without precompression. The noncubic, noninteger scaling, however, was not observed in the computational simulations by Lee {\it et al.} \cite{Lee.2015b}. While this likely occurs due to the choice of a steep repulsive potential leading to a lack of a particle overlap and, therefore, a lack of contacts with a Hertzian repulsion. These two studies potentially demonstrate the importance of the ratio of the Hertzian repulsion and particle softness relative to other interactions. For integer scaling to be observed, the applied strain must be much smaller than the precompression strain. In the present system, this limit is below the experimentally observable range. Soft particles, which have a higher preloading, should shift the limit on imposed strain to a value above the displacement limit.  Computational models might be useful in helping to predict the connection between the magnitude of the Hertzian contact $A$ and the structure of the network (if any). While there has been some work linking the structure of capillary suspension networks to the plateau shear modulus~\cite{Bossler.2018, Bindgen.2020}, the link to the magnitude of the Hertzian repulsion in the third harmonic, as used here, remains unexplored. Our experimental observations of the fit parameter would serve as an important database for testing such a computational model.
Finally, while the present model can also explain the response in other particle-based systems, there may be other mechanisms to consider and falsify, such as  thixotropic structure evolution exponents \cite{Blackwell.2016} or nonaffine motion of the sample-spanning network structure~\cite{Pine.2005, Pham.2016}. Indeed, different scalings obtained for the forward and reverse ramps may hint at flow-induced forces which break the contacts leading to the different elastic scalings for the ramp down. Interestingly, $m_\mathrm{3,elastic}$ tends to be slightly higher for the ramp up for the pendular state suspensions (Figure S4) whereas it is higher for the ramp down in the capillary state suspensions (Figure S9). Since the elastic and viscous scalings are sensitive to changes to the microstructure, further experiments connecting these microstructural changes during the ramps should be explored in the future. Particle contacts can explain the majority of reports for the noninteger and noncubic scalings, including the present systems; an alternative explanation would be required to determine why such a scaling is observed in, e.g., HDPE melts.   

\section*{Supplementary Online Material} \label{SOM}
Additional figures and tables are available in the Supplementary Online Material.

\section*{Acknowledgements} \label{acknowledgement}
We would like to thank Altuglas International for providing the BS 100 PMMA beads. Furthermore, we would like to acknowledge financial support from the European Research Council under the European Union's Seventh Framework Program (FP/2007-2013) / ERC grant agreement no. 335380, Karlsruhe House of Young Scientists for the networking travel grant, which makes this collaboration possible, for the support from the Research Foundation Flanders (FWO) Odysseus Program (grant agreement no. G0H9518N). RHE acknowledges support from the US National Science Foundation under Grants No. CMMI-1463203 and No. CBET-1351342.

%\bibliographystyle{elsarticle-num}
%\bibliography{bibliographycaps}

% % insert the supplementary information
% \newpage

% \appendix
% \input{SI_2ndV2.tex}

\end{document}